\documentclass[aps,prx,amssymb,amsmath,superscriptaddress,twocolumn,10pt]{revtex4-1}
\usepackage[T1]{fontenc}
\usepackage[latin9]{inputenc}
\usepackage{amsmath}
\usepackage{graphicx}
\usepackage[monochrome]{color} 
\usepackage{longtable}
\usepackage{subfigure}
\usepackage{soul}

\begin{document}

\title{Topological phases in two-legged Heisenberg ladders with alternating interactions}

\author{Greta Ghelli}
\affiliation{Dipartimento di Fisica e Astronomia dell'Universit\`a di Bologna, I-40127 Bologna, Italy}

\author{Giuseppe Magnifico}
\affiliation{Dipartimento di Fisica e Astronomia dell'Universit\`a di Bologna, I-40127 Bologna, Italy}
\affiliation{INFN, Sezione di Bologna, I-40127 Bologna, Italy}

\author{Cristian Degli Esposti Boschi}
\affiliation{CNR-IMM, Sezione di Bologna, I-40129 Bologna, Italy }

\author{Elisa Ercolessi}
\affiliation{Dipartimento di Fisica e Astronomia dell'Universit\`a di Bologna, I-40127 Bologna, Italy}
\affiliation{INFN, Sezione di Bologna, I-40127 Bologna, Italy}

\begin{abstract}
We analyze the possible existence of topological phases in two-legged spin ladders considering a staggered interaction in both chains. When the staggered interaction in one chain is shifted by one site with respect to the other chain, the model can be mapped, in the continuum limit, into a non linear sigma model NL$\sigma$M plus a topological term which is nonvanishing \textcolor{red}{when} the number of legs is two. This implies the existence of a critical point which distinguishes two phases. We perform a numerical analysis of energy levels, parity and string non-local order parameters, correlation functions between $x,y,z$ components of spins at the edges of an open ladder, the degeneracy of the entanglement spectrum and the entanglement entropy in order to characterize these two different phases. We identify one phase with a Mott insulator and the other one with a Haldane insulator.  
\end{abstract}

\maketitle

\section{Introduction }
\label{sec:introduzione}

Heisenberg spin model and its anisotropic variants (for a review see \cite{DEM} and references therein) represent an ideal playground for the description of quantum phases of matter \cite{Sadchev} with magnetic degrees of freedom. In the special case of one-dimensional models, these systems exhibit both gapless and gapped phases, with para-, ferro- or antiferro-magnetic correlations, and have been extensively used as a benchmark to develop new analytical or perturbative techniques. In particular, the spin $s=1/2$ $SU(2)$ antiferro-magnetic (AFM) Heisenberg chain, which  was known to be exactly solvable \cite{Baxter} and to correspond to a critical model (with a non-magnetic ground state with short range correlations only), has been assumed for decades as a paradigm. Thus, it came to a surprise when, in 1982, Haldane \cite{Haldane1,Haldane2} argued that the spin $s=1$ chain is instead gapped. Indeed, Haldane's conjecture states that there is a substantial difference between half-integer and integer spin chains. Such different behaviour can be explained in a semi-classical approach which makes use of spin coherent states \cite{KlauderSkagerstam} by mapping the model, in the continuum and low-energy limit, to an effective $O(3)$ non-linear sigma model (NL$\sigma$M) \cite{Haldane1,Haldane2,Affleck2} plus a topological term \cite{Rajaraman,Morandi}, whose coefficient $\theta$ is proportional to the value of the spin $s$. For half-integer spin, the topological term is an odd multiple of $\pi$ and thus weights the different topological sectors with an alternating sign, giving rise to a massless spectrum \cite{ReadSachdev}. On the contrary, the topological term is a multiple of $2\pi$ and thus is uneffective for integer spin, resulting in a pure $O(3)$ NL$\sigma$M which is a massive theory characterized by a finite correlation length \cite{Polyakof}. 

Actually, Haldane's argument relied on the assumption that the spin was large, but, mainly based on numerical checks, it was expected that its conclusions could be extended also to lower spins, down to $s = 1$. In 1987, Affleck, Kennedy, Lieb and Tasaki  \cite{AffleckKennedyLiebTasaki1} introduced the so-called AKLT model, for which the exact ground state and the existence of the Haldane gap was obtained analytically. This was just the first example of a whole new class of models exhibiting gapped phases which were soon proved to be characterized by \cite{AffleckKennedyLiebTasaki2,KennedyTasaki} hidden symmetries and non-local order parameters (NLOP). Similarly to what happens for the classical $XY$ model and its BKT transition \cite{Berezinskii,KosterliztThoulesse}, it was known that for all these quantum Hamiltonians, containing short-range interactions only, the Mermin-Wagner theorem \cite{Hohenberg,MerminWagner} would prevent the breaking of any continuous ($SU(2)$ or $U(1)$) symmetry, yielding instances of what we now call symmetry protected  topological (SPT) order, in which the standard framework of the Ginzburg-Landau theory \cite{Landau} is not applicable. This is similar to what was then recognized to happen in a variety of models with fermions \cite{GuWen,SchnyderRyuFurusakiLudwig,Kitaev,Wen,FidkowskiKitaev}, including topological insulators and superconductors \cite{HK,QZ}. Recently, it has been pointed out \cite{MontorsiRoncaglia,BarbieroMontorsiRoncaglia} that a suitable class of NLOPs might provide a complete classification for fermionic models as well \cite{MontorsiDolciniIottiRossi}, at least in the weak coupling regime, as long as they might be dealt within a bozonization approach and mapped to a sine-Gordon theory \cite{Giamarchi}.

The Heisenberg model in the two dimensional case is very different \cite{Haldane3,FradkinStone,IoffeLarkin,DombreRead,AngelucciJug}: the topological term is absent whatever the spin is and whatever the topology of the bipartite lattice is. To have a behaviour similar to what happens in one dimension, one should examine quasi-bidimensional models such as coupled chains, i.e. spin ladders. For Heisenberg ladders, a generalized  "even-odd conjecture" was put forward \cite{RiceGolapanSigrist}, according to which ladders with integer spin are gapped, while ladders with half-integer spin are gapless if the number of legs is odd and gapped if the number of legs is even, a fact strongly supported by numerical checks \cite{DagottoRieraScalapino,Dagotto,JohnstonJohnstonGoshornJacobsen,HiroiAzumaTakanoBando,Azuma,BarnesRiera,DagottoRice}. 
The  existence of topological features as the cause of this different behaviour in spin ladders was first investigated in \cite{Khveshchenko} and in \cite{Senechal}. \textcolor{red}{Antiferromagnetic spin ladders were studied in \cite{SheltonNersesyanTsvelik} by using bosonization techniques, while, } following the original Haldane's mapping, Dell'Arringa et al. in \cite{ArringaErcolessiMorandiPieriRoncaglia} and Sierra in \cite{Sierra1,Sierra2} mapped the Heisenberg Hamiltonian of a spin ladder into a NL$\sigma$M which contains a topological term whose coefficient $\theta$ is proportional to both spin and number of legs, proving the above mentioned conjecture. 

It is also known that a way to control the coefficient in front of the topological term in an independent way with respect to the value of the spin is by introducing alternating interactions. This has been considered for example in \cite{VenutiBoschiErcolessiOrtolaniMorandiPasiniRoncaglia,Sierra2,Sato} for the one-dimensional chain and in \cite{MartinShankarSierra} for ladders, showing that in all cases a critical point is expected for the value of the parameter controlling  \textcolor{red}{the alternated interactions} which yields a coefficient of the topological term $\theta_c=\pi$. Such critical point separates two different gapped phases whose properties we want to investigate in this paper. In particular we will show that it is indeed the presence of the topological term in the NL$\sigma$M of the corresponding effective continuum field theories that controls the emergence of a phase with an SPT order, which shows up for $\theta > \theta_c$. We remark that a similar connection between topological terms in the continuum effective Lagrangian and the appearance of a SPT phase has been found also in a one-dimensional fermionic system which aims at describing a generalization of the lattice version of Schwinger model for 1+1-dimensional quantum electrodynamics \cite{Bermudez1,Bermudez2}.

The paper is organized as follows.\\
In Sect.\ref{sec:analitico} we present our Hamiltonian of a two-legged spin ladder, with alternated Heisenberg interactions along each chain. Following \cite{ArringaErcolessiMorandiPieriRoncaglia,VenutiBoschiErcolessiOrtolaniMorandiPasiniRoncaglia} we sketch the derivation of its continuum limit low energy effective theory, finding a NL$\sigma$M plus a topological term which is nonvanishing even if the number of legs is two. This allows us to verify the results of \cite{MartinShankarSierra,MartinDukelskySierra}, which predict a critical point for a certain value of the parameter which controls the alternation.\\
In Sect.\ref{sec:numerico} we start the numerical analysis, which confirms the existence of such a critical point, separating two different gapped phases. One of these phases (for $\theta>\theta_c=\pi$) is characterized by a set of zero modes degenerate with the ground state. On the contrary to what happens in others Heisenberg ladder models characterized by some non-topological zero energy modes \cite{NeilRobinson}, in our case we prove that the phase with zero modes encodes an SPT order investigating NLOP, namely parity and string non-local order parameters. Furthermore, in the phase with zero modes, we also check that spin correlation functions between spins at the ends of the ladder are different from zero, supporting the idea that we are in presence of edge states. Finally, we perform an analysis of the entanglement entropy and of the degeneracy of the entanglement spectrum, showing that the latter has indeed an even degeneracy in the supposed SPT phase. These results, concerning the critical point and the topological nature of one of the two phases, are also consistent with \cite{NataliaChepiga}, where our model is studied through a Berry phase investigation.\\
We finally summarize our conclusions in Sect.\ref{sec:conclusione}. Following the classification suggested in \cite{MontorsiRoncaglia,BarbieroMontorsiRoncaglia,MontorsiDolciniIottiRossi}, we can say that our numerical analysis allows us to identify the region for $\theta<\theta_c$ with a Mott insulator-like phase, with no edge states, nonvanishing value of the parity non-local order parameter and an odd degeneracy of the entanglement spectrum, while the region with $\theta>\theta_c$ with a Haldane insulator-like phase, characterized by edge states, a nonvanishing value of the string non-local order parameter and an even degeneracy of the entanglement spectrum.

\section{The model and analytical predictions}
\label{sec:analitico}

\begin{figure}
\centering
\includegraphics[width=0.5\textwidth]{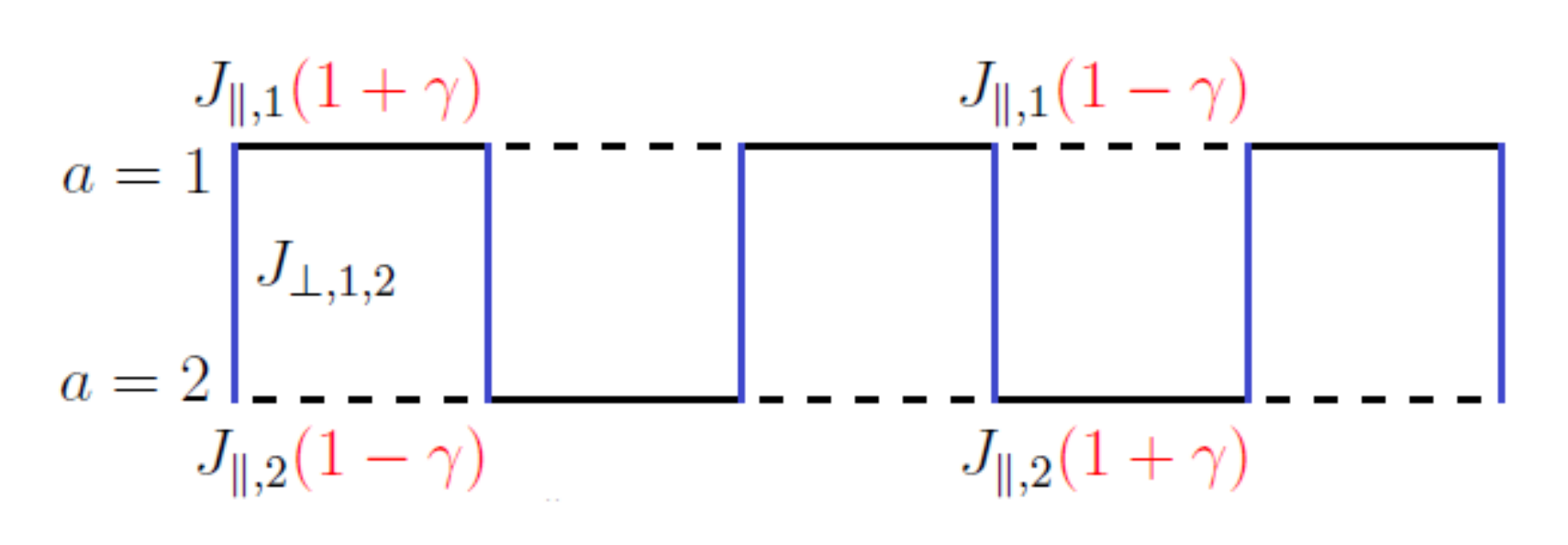}
\caption{ \textcolor{red}{Representation of model B corresponding to Hamiltonian (\ref{2}), for a chain of six sites. In this figure, the coupling constants $J_{\parallel,a}$ ($a=1,2$) and $J_{\perp,1,2}$ are indicated: strong and weak inter-chain bonds are represented by black continuous and dashed lines, while bonds between chains are in blue.}}
\label{fig:schemaModello} 
\end{figure}

We focus on two-legged spin ladders with staggered interactions along each chain. We have the choice to put the alternation in two possible different ways:
\begin{itemize}
\item A. In the same way on both chains, thus forming a columnar pattern of strong and weak bonds. The Hamiltonian reads:
\begin{align}\label{1}
H & =\sum_{a=1,2} \sum_{k=1}^{N} J_{\parallel,a} ( 1+{(-1)}^k \gamma ) 
\textcolor{red}{ \vec{S}_{k,a} \cdot \vec{S}_{k+1,a} } + \nonumber \\
  & \textcolor{red}{ + \sum_{k=1}^{N} J_{\perp,1,2} 
 \vec{S}_{k,1} \cdot \vec{S}_{k,2} } 
\end{align}
where \textcolor{red}{$S_{k,a}^{\alpha}$  ($\alpha=x,y,z$) are the components of the spin $1/2$ operator; the} index \textcolor{red}{$a=1,2$} labels the chains, while the index \textcolor{red}{$k=1,...,N$} the sites along each chain.
We will show that the topological term is zero in this case. \textcolor{red}{This model was analytically analyzed in \cite{TotsukaSuzuki}}.
\item B. In the opposite way in one chain with respect to that of the second chain, yielding a staggered pattern of strong and weak bonds, \textcolor{red}{as shown in  Fig.\ref{fig:schemaModello}. In this case, the Hamiltonian is}:
\begin{align}\label{2}
H & =\sum_{k=1}^{N} J_{\parallel,1} ( 1+{(-1)}^{k-1} \gamma ) 
\textcolor{red}{ \vec{S}_{k,1} \cdot \vec{S}_{k+1,1} } + \nonumber \\
  & +\sum_{k=1}^{N} J_{\parallel,2} ( 1+{(-1)}^{k} \gamma ) 
\textcolor{red}{ \vec{S}_{k,2} \cdot \vec{S}_{k+1,2} } + \nonumber \\
  &\textcolor{red}{ + \sum_{k=1}^{N} J_{\perp,1,2} 
 \vec{S}_{k,1} \cdot \vec{S}_{k,2} }. 
\end{align}
An equivalent situation is obtained by exchanging the role of the two chains. This is the case we will concentrate on, since a nonvanishing topological term will be present. 
\end{itemize}
We assume that the coupling constants $J_{\parallel,1}$, $J_{\parallel,2}$, $\textcolor{red}{{J}_{\perp,1,2}}$ are all positive, so the classical minimum of the Hamiltonian is antiferromagnetically ordered, and we will work in the range $-1\leq \gamma\leq 1$.\\ 
\\
The partition function of both models (\ref{1}) and (\ref{2}) can be expressed using a path integral representation 
\begin{equation}\label{3}
Z=\int \mathcal{D}\hat{\Omega} \exp \left( is \sum_{\textcolor{red}{{k,a}}} \omega[{\hat{\Omega}}_{\textcolor{red}{{k,a}}}(\tau)]-\int_0^{\beta} d\tau H(\tau) \right)
\end{equation}
with spin coherent states \cite{KlauderSkagerstam}, obtained by replacing the spin operators $\textcolor{red}{\vec{S}_{k,a}}$ with the classical variables $s\textcolor{red}{{\hat{\Omega}}_{k,a}(\tau)}$. In (\ref{3}) the first term is the Berry phase contribution, which arises as a consequence of the nonvanishing overlap between coherent states at consecutive times \cite{Auerbach} and represents the area bounded by the trajectory parameterized by ${\hat{\Omega}}(\tau)$ on the $S^2$ sphere \cite{Affleck2,Fradkin}. To calculate the action that appears in the phase of the exponential, we will assume Haldane's mapping \cite{Haldane3} and follow \cite{ArringaErcolessiMorandiPieriRoncaglia} to specialize it to the case of spin ladders, by taking:
\begin{equation}\label{4}
\textcolor{red}{{\hat{\Omega}}_{k,a}(\tau)} = (-1)^{a+\textcolor{red}{k}} \hat{\phi}(\textcolor{red}{k},\tau) {\left( 1- 
\frac{{|\textbf{l}_a(\textcolor{red}{k},\tau)|}^2}{{s}^2} \right)}^{\frac{1}{2}} + \frac{\textbf{l}_a(\textcolor{red}{k},\tau)}{s}
\end{equation}
where the spin coherent field has been written in terms of a slow-varying field $\hat{\phi}(\textcolor{red}{k},\tau)$ of unit norm, which is weighted by a staggered factor $(-1)^{\textcolor{red}{k}+a}$, and of uniform fluctuations $\textbf{l}_a(\textcolor{red}{k},\tau)$ which are assumed to be small, $|\textbf{l}_a(\textcolor{red}{k},\tau)|/s \ll 1$. This allows to expand all expressions up to quadratic order in the latter field which can then be integrated out. Notice that we take $\hat{\phi}(\textcolor{red}{k},\tau)$ not changing along a rung,  meaning that the staggered spin-spin correlation lenght $\xi$ is greater with respect to the total width of the ladder $n_l a$, a fact which is confirmed numerically \cite{GreveBirgeneauWiese,WhiteNoackScalapino}. 
Here we do not give further details of the calculations, that can be found in \cite{ThesisGhelli}. For both cases $A,B$ above, we find a partition function 
\begin{equation}\label{15}
Z=\int \mathcal{D}\hat{\phi} \exp \left( -\int dx d\tau \mathcal{L}(x,\tau) \right)
\end{equation}
where the Lagrangian density $\mathcal{L}(x,\tau)$ is written as that of a NL$\sigma$M with a topological term:
\begin{align}\label{17}
\mathcal{L}(x,\tau) & = \frac{1}{2g} \left( \frac{1}{v_s} {\dot{\hat{\phi}}}^2(x,\tau) + v_s {\hat{\phi '}}^2(x,\tau) \right) + \nonumber \\
                    & + \frac{i\theta}{4\pi} \hat{\phi '}(x,\tau) \cdot \left( \hat{\phi}(x,\tau) \times \dot{\hat{\phi}}(x,\tau)\right).
\end{align}
where
\begin{equation}\label{23}  \nonumber
\frac{1}{g}=\sqrt{ \sum_{d,b} L^{-1}_{d,b} \left( -4s^2{\gamma}^2\sum_{d,b} \alpha_{A,B} J_{\parallel,d} L^{-1}_{d,b} J_{\parallel,b} + s^2\sum_a J_{\parallel,a} \right) } , 
\end{equation}
\begin{equation}\label{24} \nonumber
 v_s =  
\sqrt{ \frac{\left( -4s^2{\gamma}^2\sum_{d,b}  \alpha_{A,B} J_{\parallel,d} L^{-1}_{d,b}  J_{\parallel,b} + s^2\sum_a J_{\parallel,a} \right)}
{\sum_{d,b} L^{-1}_{d,b}} } , 
\end{equation}
with $\alpha_{A} =(-1)^{(d+b)}$ and $\alpha_{B}=1$ while 
\begin{align}\label{25}
& \theta_{A,B}= -4\pi s\gamma \times \nonumber \\
& \left( \frac{
\mp 2J_{\parallel,1}\left(4J_{\parallel,2} + J_{\perp,1,2}\right)
+2J_{\parallel,2}\left(4J_{\parallel,1} + J_{\perp,1,2}\right)
}{
16J_{\parallel,1}J_{\parallel,2}+4J_{\parallel,1}J_{\perp,1,2}+4J_{\parallel,2}J_{\perp,1,2}
}+ \right. \nonumber \\
& \left. +\frac{ 
\pm 2J_{\parallel,1}J_{\perp,1,2}
-2J_{\parallel,2}J_{\perp,1,2}
}{
16J_{\parallel,1}J_{\parallel,2}+4J_{\parallel,1}J_{\perp,1,2}+4J_{\parallel,2}J_{\perp,1,2}
} \right).
\end{align}
We notice that our results are consistent with those of reference \cite{ArringaErcolessiMorandiPieriRoncaglia} in absence of a staggered interaction, i.e. $\gamma=0$, which in turn are consistent with those found in \cite{Senechal} and in \cite{Sierra1,Sierra2}.\\

As anticipated before, the topological term (\ref{25}) is null for case A, yielding instead a nontrivial contribution in case B, which we will concentrate on in the following.\\
It is well known \cite{ShankarRead} that the NL$\sigma$M is gapped for all values of the coefficient of the topological term, but for $\theta = \pi$, at which one finds a quantum phase transition \cite{Sadchev}. In the next sections we will check numerically this is indeed the case and we will characterize the two phases. For simplicity, we set $J_{\parallel,1}=J_{\parallel,2}=J_{\perp,1,2}=1$ which implies that $\theta_c=\pi$ when $\gamma_c=-0.75$. This result is the same found in \cite{MartinShankarSierra, MartinDukelskySierra}. 

\section{Numerical analysis}
\label{sec:numerico}

Our numerical analysis is based on the Density Matrix Renormalization Group algorithm \cite{DMRG1,DMRG2} using \textcolor{red}{Matrix Product State} (MPS) tensor network \cite{ver,sch,TN}.

\subsection{Energy levels and critical point}

Our first purpose is to look for \textcolor{red}{the existence of} a critical point, by looking at the gap. 

\begin{figure}
\centering
\includegraphics[width=0.5\textwidth]{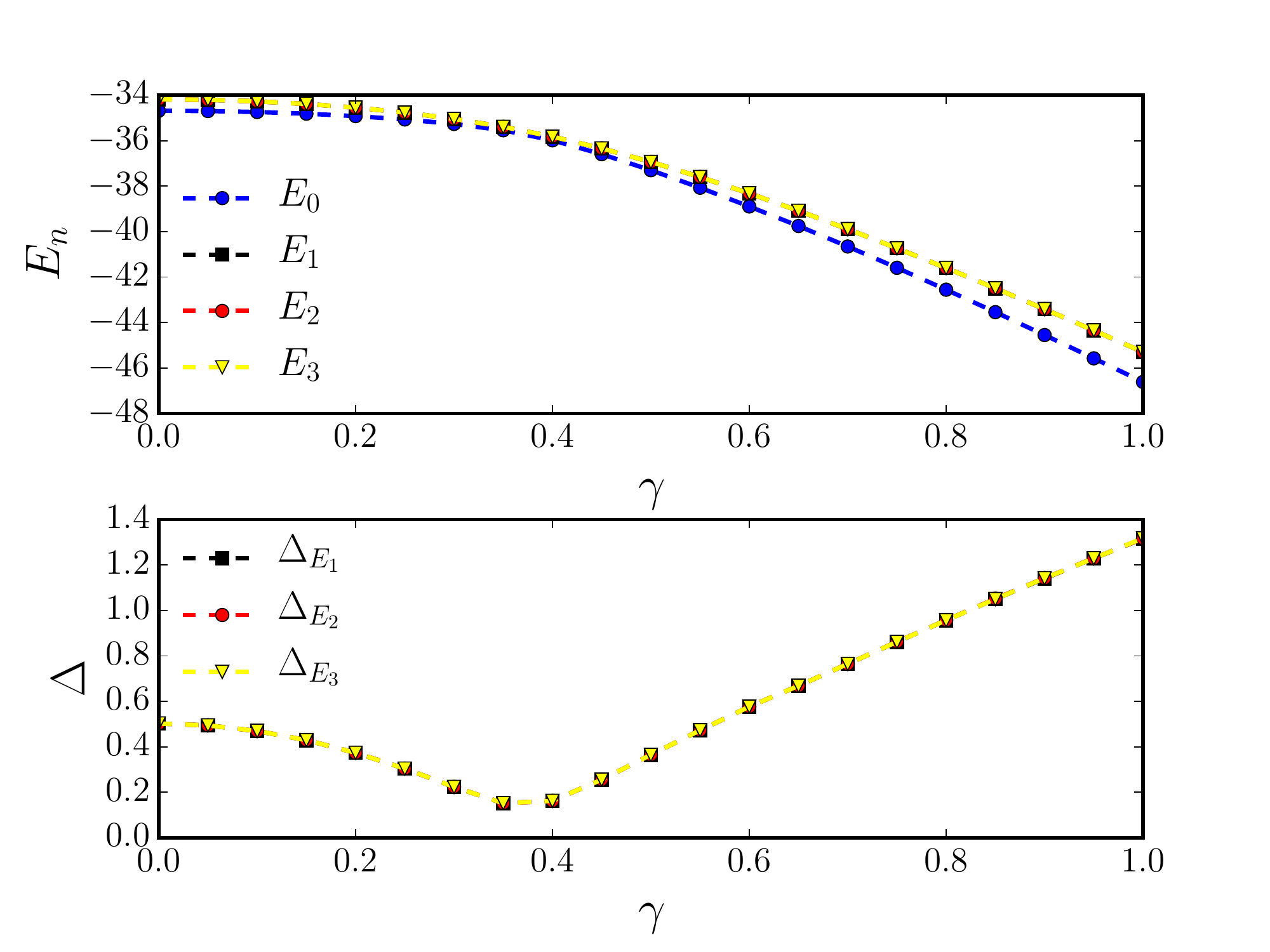}
\caption{\textcolor{red}{The ground state energy and the first three excited states as a function of the parameter $\gamma$} (top panel) and \textcolor{red}{their} gaps with respect to the ground state (bottom panel) in the case of PBC. Here $\gamma$ varies from 0 to 1 with a 0.05 step and we consider $N=30$ sites for each chain.}
\label{fig:energylevelsPBC} 
\end{figure}
Fig.\ref{fig:energylevelsPBC} shows the results for \textcolor{red}{periodic boundary conditions (PBC)}: in the top panel, the energies of the ground state (which is in the subspace \textcolor{red}{$S_{tot}^{z}=\sum_{a=1,2}\sum_{k=1}^{N} S_{k,a}^{z}=0$}) and of the first three excited states (which, because of the $SU(2)$ symmetry, are degenerate and belong to subspaces \textcolor{red}{$S_{tot}^{z}=0$}, \textcolor{red}{$S_{tot}^{z}=+1$}, \textcolor{red}{$S_{tot}^{z}=-1$}); in the bottom panel, the values of the triplet gap. \textcolor{red}{The energy of the ground state is indicated with $E_0$, while the energies of the triplet are indicated with $E_1$, $E_2$ and $E_3$. Furthermore, $\Delta_{E_1}$, $\Delta_{E_2}$ and $\Delta_{E_3}$ are their gaps with respect to the ground state.} The data have been obtained by considering $N=30$ sites on each chain for PBC and the parameter $\gamma$ varies from 0 to 1 with a 0.05 step (the model is symmetric under the inversion $\gamma\rightarrow -\gamma$).

\begin{figure}
\centering
\includegraphics[width=0.5\textwidth]{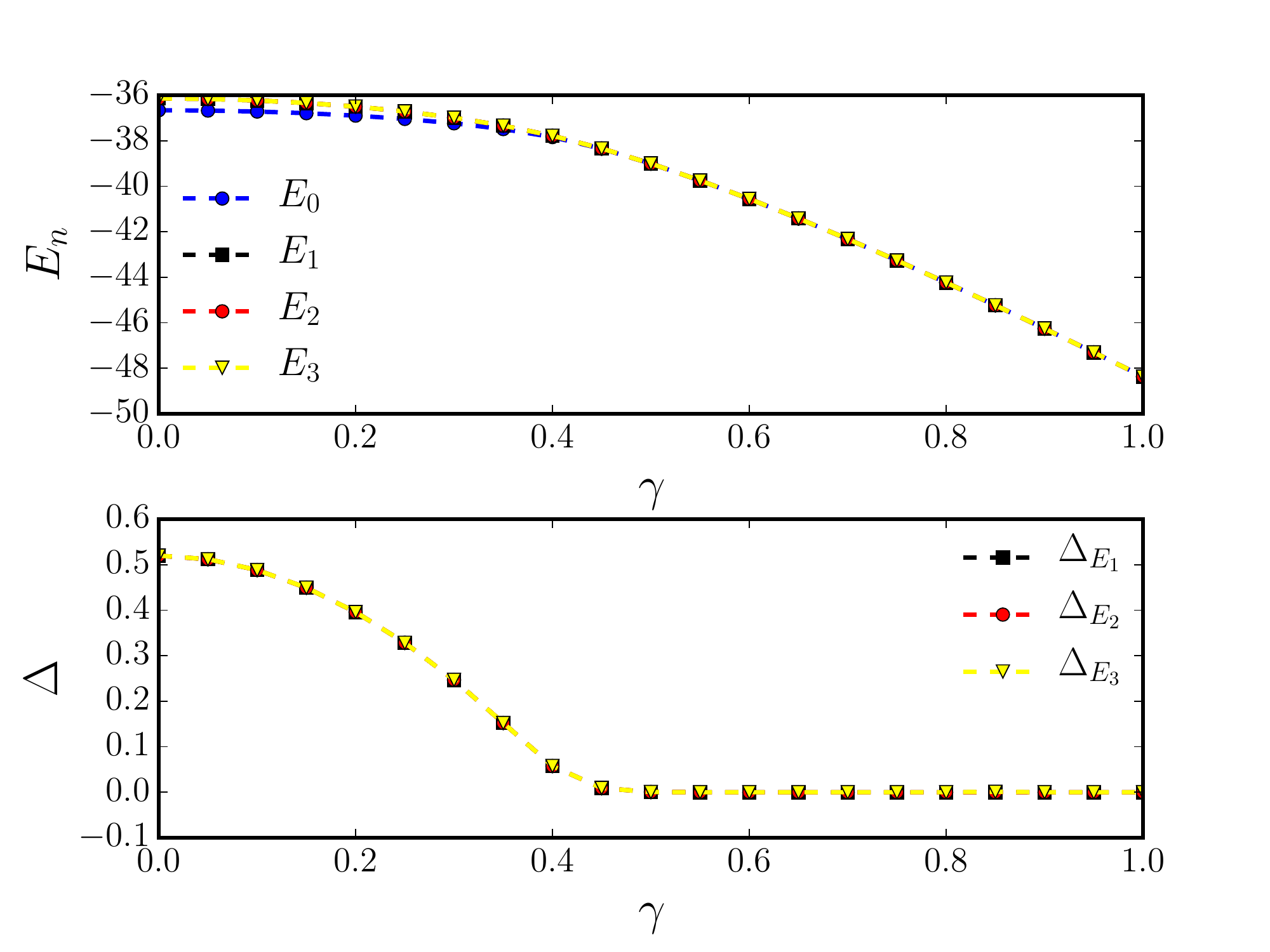}
\caption{ \textcolor{red}{The ground state energy and the first three excited states as a function of the parameter $\gamma$} (top panel) and \textcolor{red}{their} gaps with respect to the ground state (bottom panel) in the case of OBC. Here $\gamma$ varies from 0 to 1 with a 0.05 step and we consider $N=32$ sites for each chain.}
\label{fig:energylevelsOBC} 
\end{figure}

Then, we consider \textcolor{red}{open boundary conditions (OBC).} The values of the energies of the first four states and of the triplet gap are shown in Fig.\ref{fig:energylevelsOBC}, respectively in the top and in the bottom panel, \textcolor{red}{ using the same notation of Fig.\ref{fig:energylevelsPBC}. We consider $N=32$ sites on each chain and $\gamma$ varying from 0 to 1 with a 0.05 step.} 

\textcolor{red}{The data clearly show a closure of the gap for $\gamma \sim 0.35-0.4$. Even if 
the exact determination of the critical point is not our aim, we provide in the Appendix a finite-size scaling analysis for 
these values of $\gamma$, which confirms the existence of a critical point.  It is evident that $\gamma_c$ deviates a lot from the expected theoretical value $\gamma_c=0.75$. This does not come to a complete surprise, since renormalization corrections to the semiclassical analysis performed in the previous section are expected, as also remarked in \cite{MartinShankarSierra,MartinDukelskySierra}.}

We \textcolor{red}{also} notice that the triplet states are gapped in the phase for $\gamma < \gamma_c$ while they are degenerate with the ground state in the phase for $\gamma > \gamma_c$, yielding zero modes. This is signalling that the latter phase might indeed be a SPT phase, a fact that we are going now to prove. 

\subsection{Non-local order parameters}

Let us remark that our system is essentially a one-dimensional model with two species of spin, one for each chain. Using a Jordan-Wigner transformation  \cite{Auerbach,GogolinNersesyanTsvelik}, it can be interpreted as an interacting system of two fermionic species whose densities $n_a/N$ $(a=1,2)$ are separately conserved. As for the Hubbard model, we have that the total $z$-spin and total charge densities, respectively defined as $(n_1 + n_2)/N$ and  $(n_1 - n_2)/N$, are conserved. Thus, in order to characterize the two phases of the Hamiltonian (\ref{2}) which are separated by the critical point $\gamma_c$, we can follow the work \cite{MontorsiRoncaglia,BarbieroMontorsiRoncaglia,MontorsiDolciniIottiRossi} and introduce the following two types of non-local order parameters, defined in terms of the parity and the string operators:
\begin{align}\label{26}
\textcolor{red}{C_P^{\alpha}(r)=\left\langle \prod_{k=j}^{j+r-1} e^{i\pi (S_{k,1}^{\alpha}+S_{k,2}^{\alpha})
}\right\rangle} \, ,\\
\label{27}
\textcolor{red}{C_S^{\alpha}(r)=\left\langle 2S_{j,1}^{\alpha} \prod_{k=j}^{j+r-1} e^{i\pi (S_{k,1}^{\alpha}+S_{k,2}^{\alpha})} 2S_{j+r,1}^{\alpha} \right\rangle} \, .
\end{align}
Notice that in all exponentials we take the sum of the spins on both chains and put a factor $\pi$, as suggested in \cite{FazziniBeccaMontorsi,DM}. The factor 2 in $C_S^{\alpha}(r)$ is introduced because it gives the correct normalization.\\
In order to reduce as much as possible finite-size effects, we consider $1\leq r\leq \frac{N}{2}$ for PBC and $\frac{N}{4}\leq r \leq \frac{3N}{4}$ for OBC. For both $C_P^{\alpha}$ and $C_S^{\alpha}$ the initial site $j$ in and the final site $j+r$ belong to chain 1.\\ 
\begin{figure}
\centering
\includegraphics[width=0.5\textwidth]{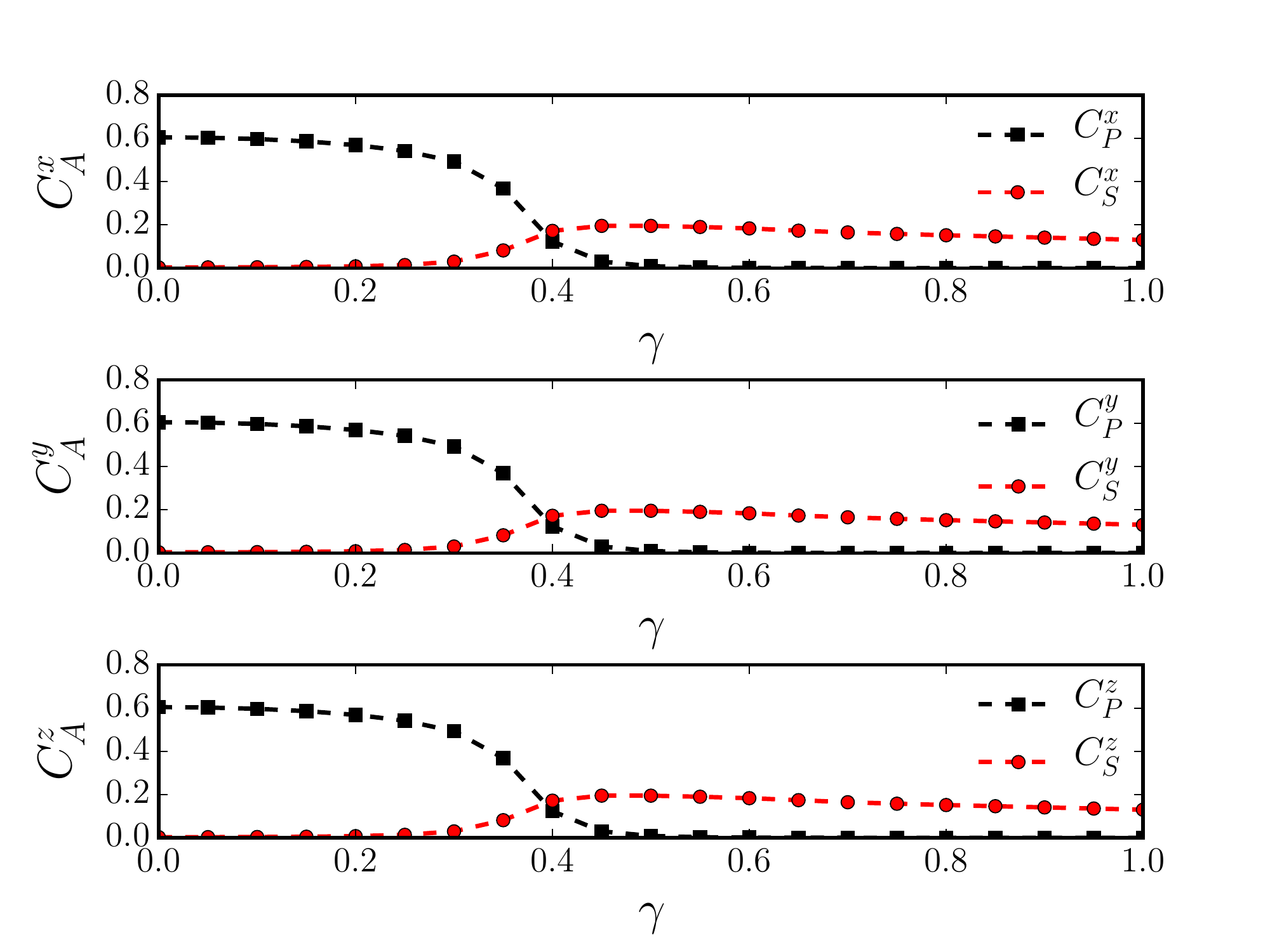}
\caption{ \textcolor{red}{Parity and string order parameters $C_{P,S}^{\alpha}$, $\alpha=x,y,z$, are shown, respectively in black and in red, as function of the parameter $\gamma$. Here we use PBC, $N=30$ sites on each chain and $\gamma$ varies from 0 to 1 with a 0.05 step.}}
\label{fig:NLOPwithPBC} 
\end{figure}
\begin{figure}
\centering
\includegraphics[width=0.5\textwidth]{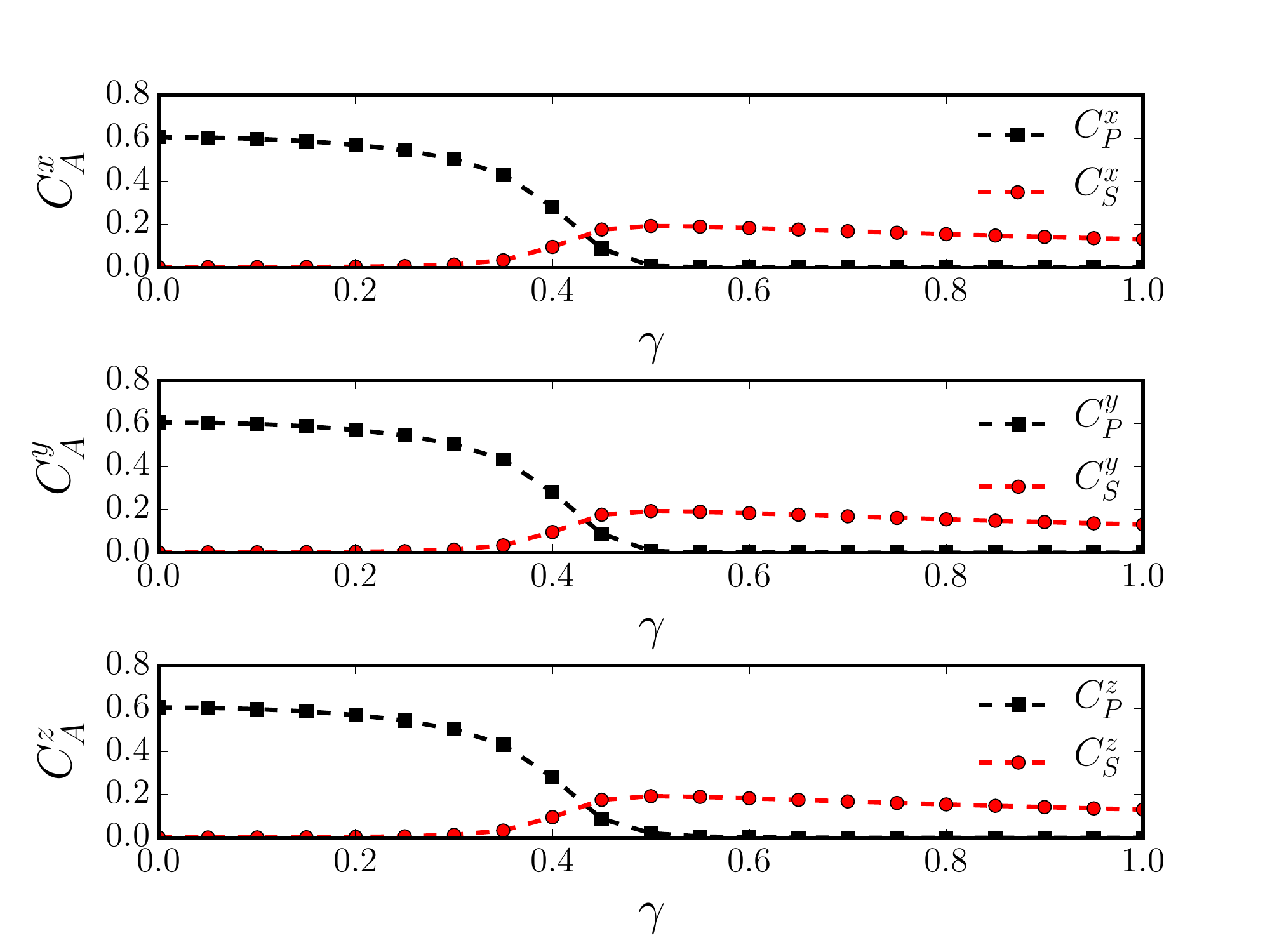}
\caption{ \textcolor{red}{Parity and string order parameters $C_{P,S}^{\alpha}$, $\alpha=x,y,z$, are shown, respectively in black and in red, as function of the parameter $\gamma$. Here we use OBC, $N=32$ sites on each chain and $\gamma$ varies from 0 to 1 with a 0.05 step.}}
\label{fig:NLOPwithOBC} 
\end{figure}
The behaviour of $C_P^{\alpha}$ (black line) and $C_S^{\alpha}$ (red line) are given in Fig.\ref{fig:NLOPwithPBC} for PBC and in Fig.\ref{fig:NLOPwithOBC} for OBC, with a chain of $N=30$ sites for PBC and $N=32$ for OBC ($\gamma$ always varies from 0 to 1 with a 0.05 step).\\
We clearly see that the $SU(2)$ symmetry is respected, so that the $x,y,z$ components of all parameters look the same. Both with PBC and OBC, the parity  and the string operator have a dual behaviour in the two phases, with $C_P^{\alpha}$ nonvanishing for $\gamma < \gamma_c$ and  $C_S^{\alpha}$ different from zero for $\gamma > \gamma_c$. 

\subsection{Correlation functions and edge states}

\begin{figure}
\centering
{\includegraphics[width=0.5\textwidth]{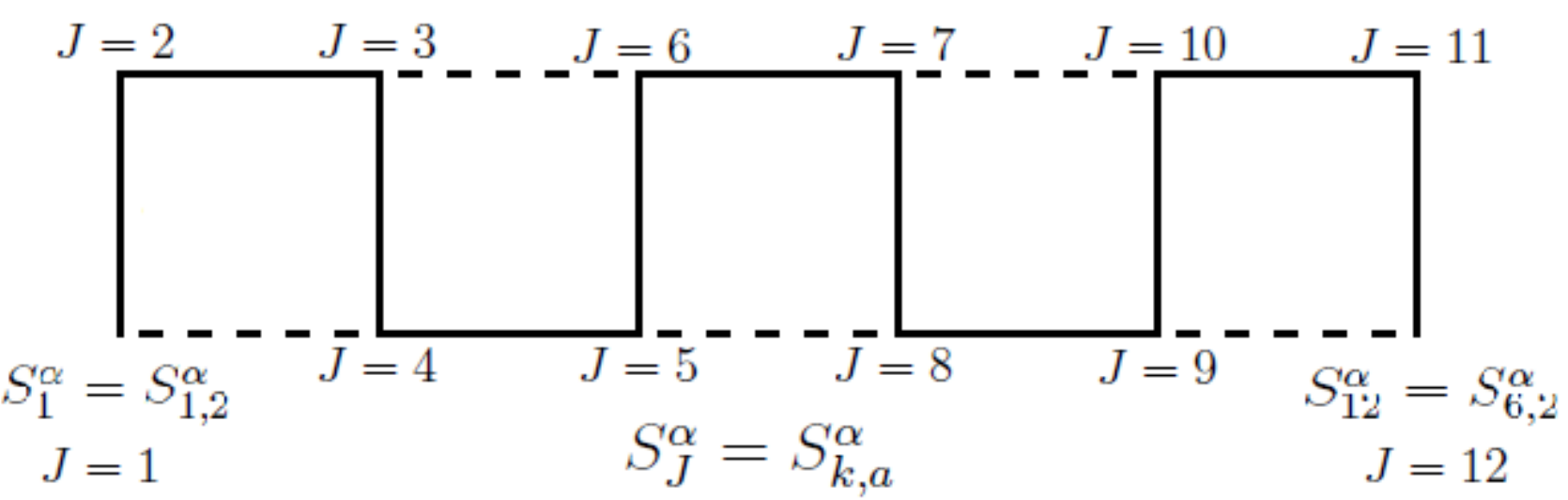}}
\hspace{2mm}
{\includegraphics[width=0.5\textwidth]{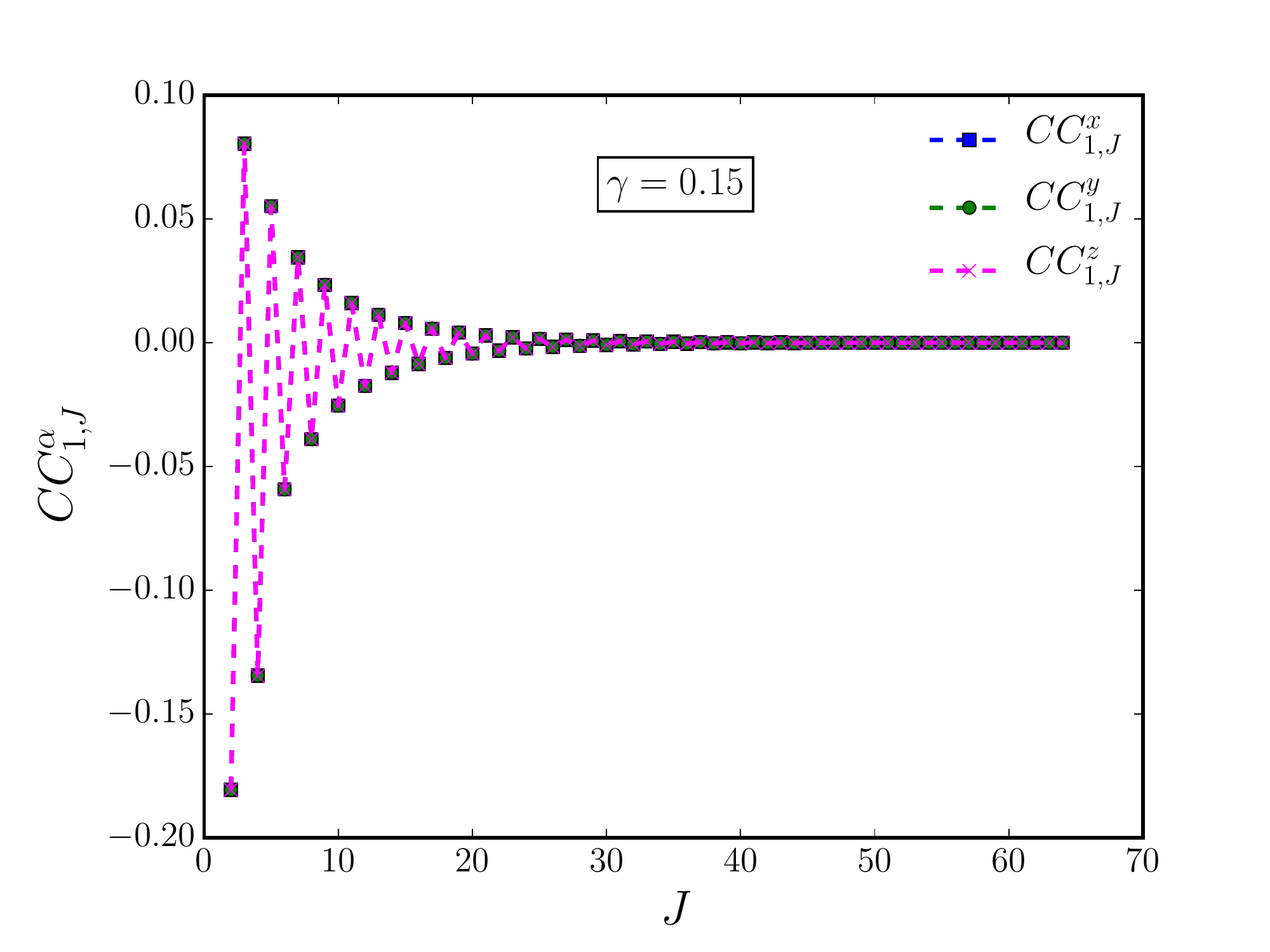}}
\caption{ \textcolor{red}{Top panel:  a ladder with $N=6$ showing the new way of labeling sites using the index $J$, by following a snake path.  Bottom panel: ground state spin correlations $CC^{\alpha}_{1,J}$ ($\alpha=x,y,z$), obtained with OBC, $\gamma=0.15$ and $N=32$ sites on each chain, i.e. $64$ sites for the ladder. Correlations are calculated between the first spin ($J=1$) of the ladder kept fixed and each of the other spins of the ladder, until the last one ($J=2N=64$).}}
\label{fig:Corr015alternata}
\end{figure}
\begin{figure}
\centering
\includegraphics[width=0.5\textwidth]{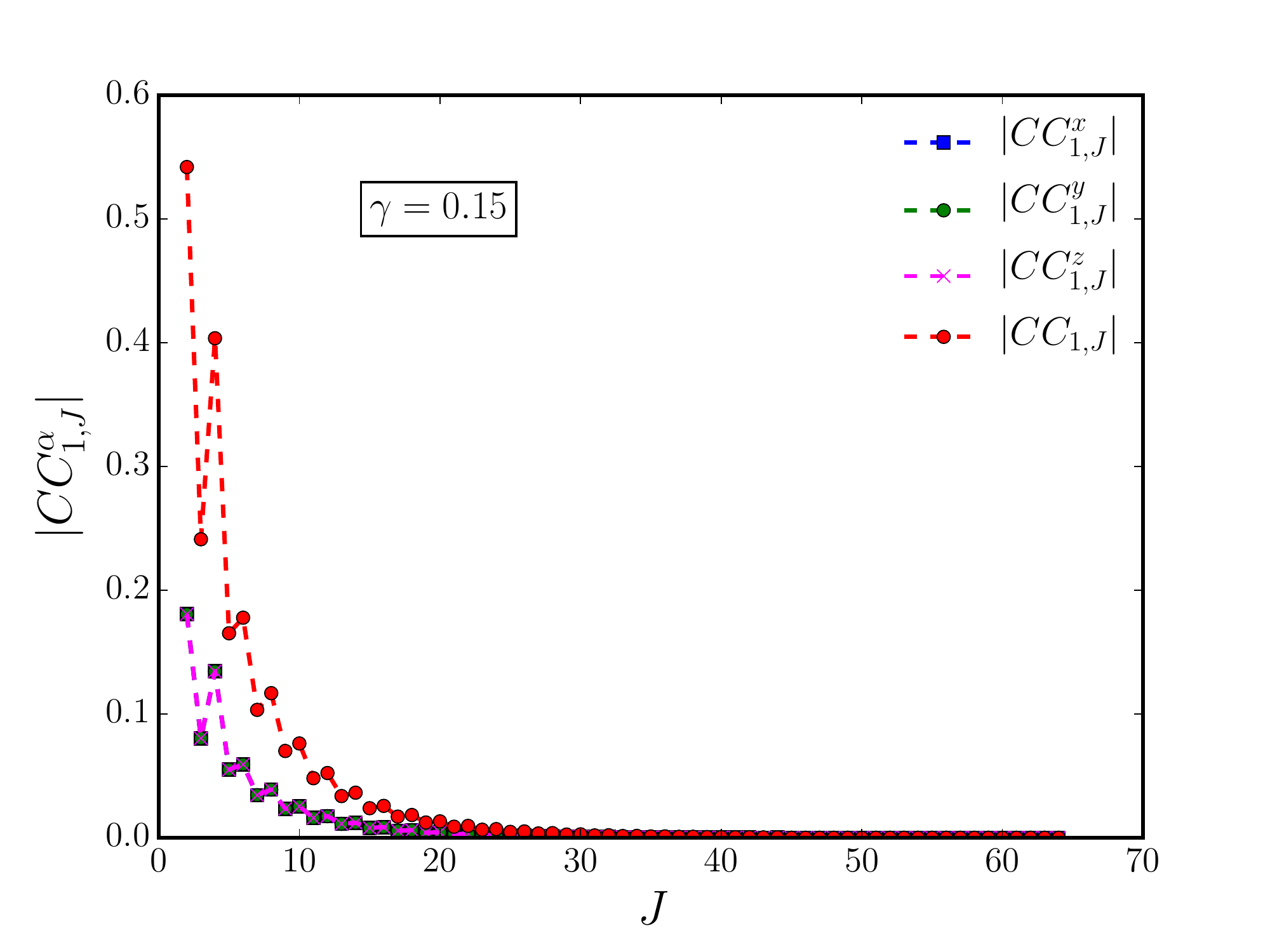}
\caption{\textcolor{red}{Absolute value of ground state spin correlations $|CC^{\alpha}_{1,J}|$ ($\alpha=x,y,z$) and  $|CC_{1,J}|$, for OBC and $N=32$ sites on each chain, i.e. $64$ sites for the ladder. Correlations are calculated between the first spin ($J=1$) of the ladder kept fixed and each of the other spins of the ladder, until the last one ($J=64$), in the trivial phase for $\gamma=0.15$.}}
\label{fig:Corr015} 
\end{figure}
\begin{figure}
\centering
\includegraphics[width=0.5\textwidth]{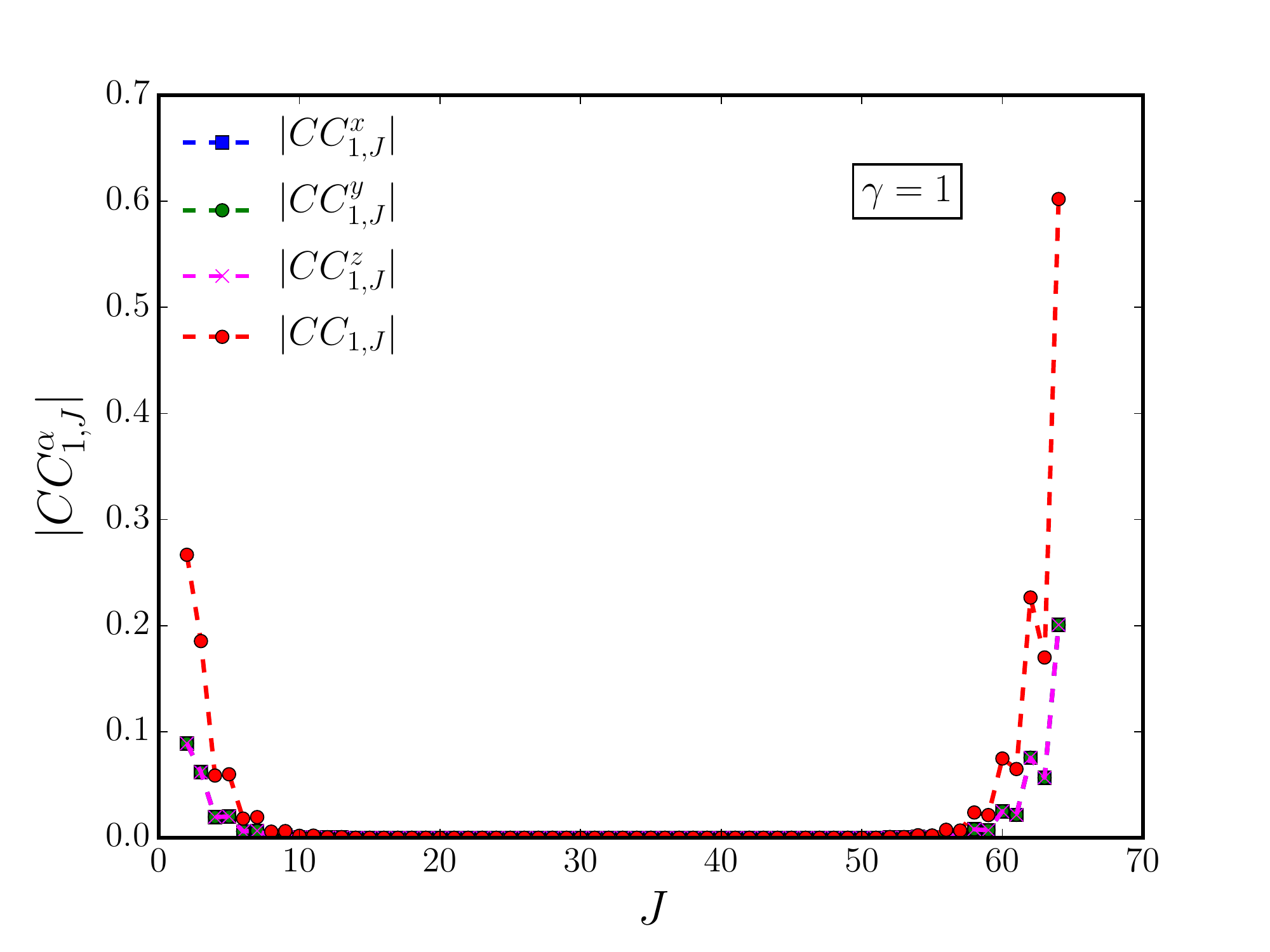}
\caption{\textcolor{red}{Absolute value of ground state spin correlations $|CC^{\alpha}_{1,J}|$ ($\alpha=x,y,z$) and  $|CC_{1,J}|$, for OBC and $N=32$ sites on each chain, i.e. $64$ sites for the ladder. Correlations  are calculated between the first spin ($J=1$) of the ladder kept fixed and each of the other spins of the ladder, until the last one ($J=64$), in the topological phase for $\gamma=1$.}}
\label{fig:Corr1} 
\end{figure}
In this subsection we investigate spin correlation functions between the first spin of the ladder kept fixed (i.e. the first spin of the second chain) and each of the others spins of the ladder until the last one (i.e. the last spin of the second chain), when considering OBC. \textcolor{red}{To simplify the notation, we adopt a new label to number the spins along the ladder: $S_J^{\alpha}$, with $J=1,\cdots,2N$, following a snake path, as shown in  the top panel of Fig.\ref{fig:Corr015alternata}. By using this new label, spin correlation functions can be expressed as follows:} 
\begin{align}\label{primadi10A}
\textcolor{red}{{CC}^{\alpha}_{1,J}=<S_1^{\alpha} S_J^{\alpha}>}
\end{align}
\begin{align}\label{primadi10B}
\textcolor{red}{{CC}_{1,J}=\sum_{\alpha} {CC}^{\alpha}_{1,J} = <\vec{S}_1 \cdot \vec{S}_J>}
\end{align}
\textcolor{red}{where $\alpha=x,y,z$ and $J=1,..,2N$.}\\
Indeed, being in a gapped phase, we expect them to decay in an exponential way for the trivial $\gamma< \gamma_c$ case, while in the supposed topological case $\gamma> \gamma_c$ they should still decay in the bulk but have a non-zero value between the first and the last spins of the ladder, signalling the appearance of edge states. 

We first notice that the $SU(2)$ symmetry, which implies that correlation functions are identical along all three directions, is respected. Also, as expected, there is a short-range AFM order, which is evident from the staggered behaviour of the correlation functions, as shown for example in 
Fig.\ref{fig:Corr015alternata}. Their absolute value is plotted in  Fig.\ref{fig:Corr015} for the trivial case ($\gamma=0.15$) and in Fig.\ref{fig:Corr1} for the topological phase ($\gamma=1$). In the latter case, \textcolor{red}{in order to sort out the ground state living in the spin zero sector, we perform the numerical simulations by adding an interaction with a small magnetic field, $\mu {\left( \sum_{J=1}^{64} \vec{S}_J \right)}^2$}, with $\mu=10^{-3}$, to the Hamiltonian (\ref{2}). 

From Fig.\ref{fig:Corr1} we clearly see that, for $\gamma> \gamma_c$, there is a strong correlation between the first and the last spin of the ladder, which we interpret as the emergence of zero modes made up of two entangled spins at the edges. Following the work \cite{VenutiBoschiRoncaglia} which characterizes long-distance entanglement in spin systems, we can quantify the degree of entanglement carried by such edge states by means of the concurrence between the first spin of the ladder \textcolor{red}{($J=1$)}  and each of the other spins of the ladder until the last one \textcolor{red}{($J=64$)}. Having an $SU(2)$ symmetry, the concurrence can be computed as:
\begin{equation}\label{28}
\textcolor{red}{C_{1,J}=\frac{1}{2}\max \left(0,-1-12 \, \, CC^{z}_{1,J} \right)}.
\end{equation}
We find that 
\begin{equation}\label{29}
\textcolor{red}{\sum_J {\left( C_{1,J} \right)}^2=0.49707302}
\end{equation}
where the major contribution (of about 99.8\%) is given by the case where \textcolor{red}{$J$} corresponds to the last spin of the ladder. We also note that this sum is smaller then 1, indicating that the system carries also a certain degree of multi-partite entanglement \cite{OsborneVerstraete} (indeed due to rotational symmetry, all the single-site magnetizations vanish and the reduced density matrix describes a maximally entangled state of one spin with all the others).

\subsection{Entanglement entropy and entanglement spectrum}

\begin{figure}
\includegraphics[width=0.5\textwidth]{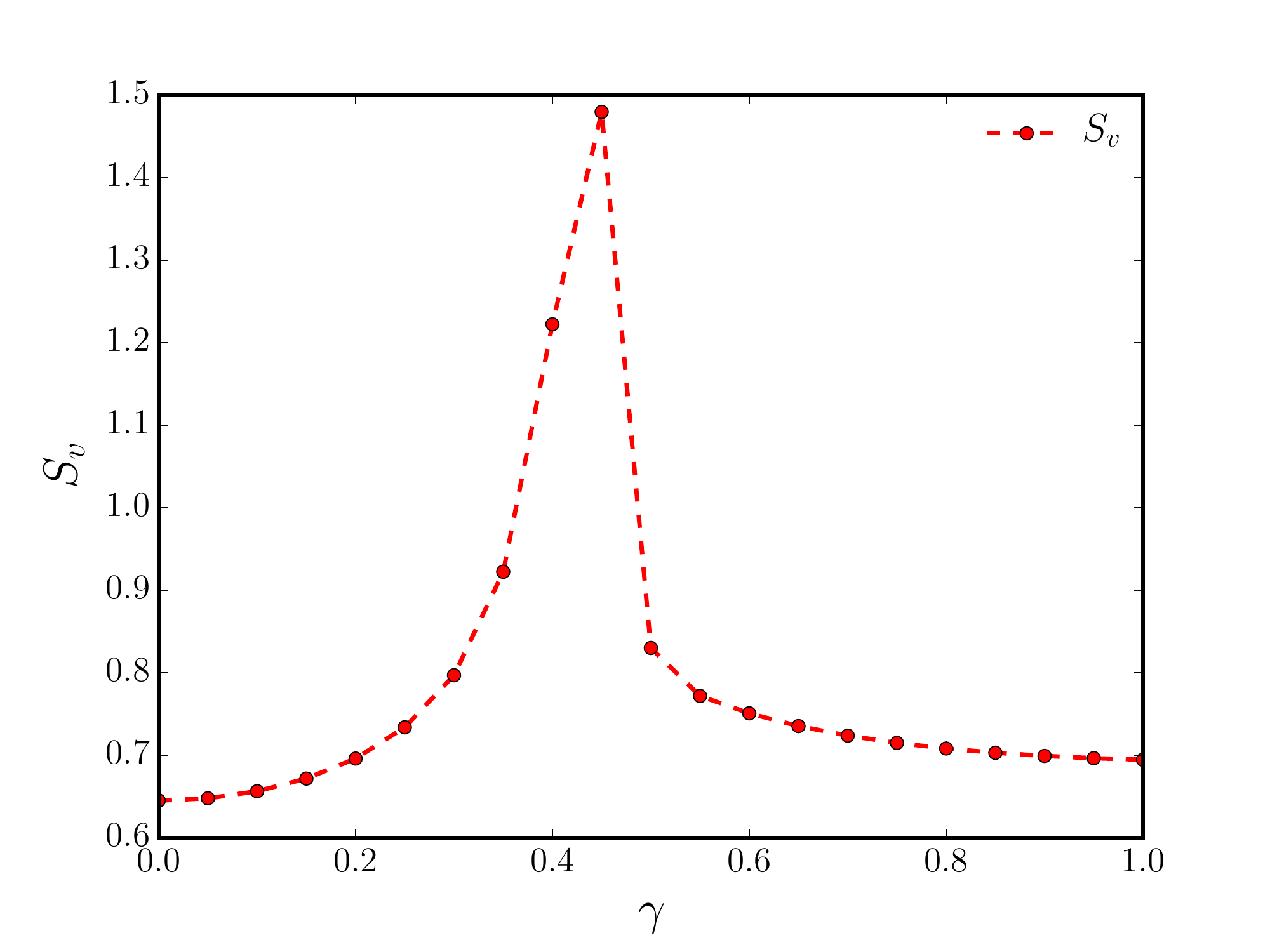}
\caption{Representation of entanglement entropy $S_v$ as function of $\gamma$ which goes from 0 to 1 with a 0.05 step. We use OBC and $N=32$ sites on each chain.}
\label{fig:Entropia} 
\end{figure}
\begin{figure}
\centering
\includegraphics[width=0.5\textwidth]{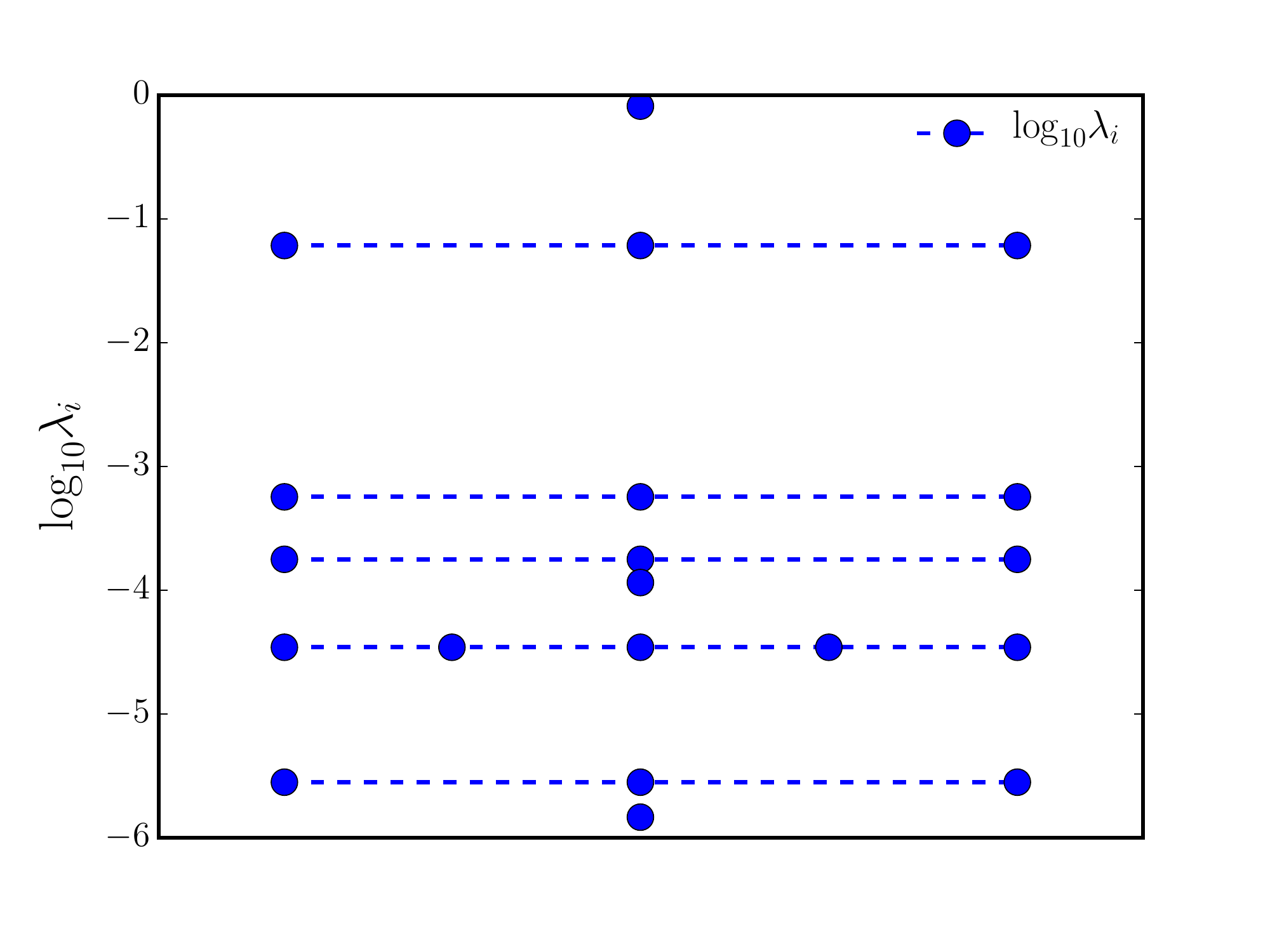}
\caption{ Representation of the the logarithm of the eigenvalues $\lambda_i$ of the reduced density matrix of the system at $\gamma=0.2$. We use OBC and $N=32$ sites on each chain. For each value of $\lambda_i$, it is possible to see horizontally the corresponding degeneration given by the number of blue dots.}
\label{fig:Autovalori02} 
\end{figure}
\begin{figure}
\centering
\includegraphics[width=0.5\textwidth]{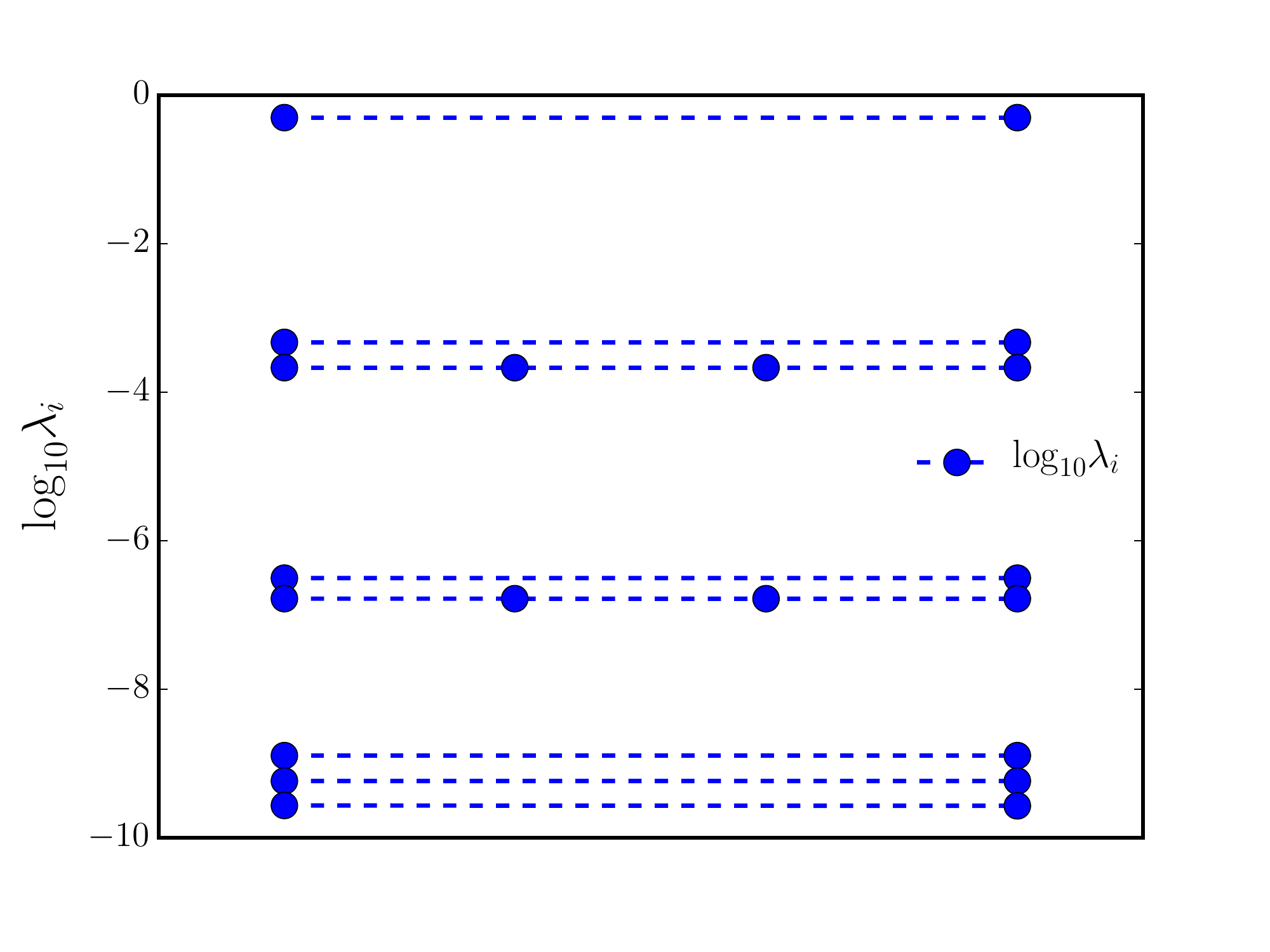}
\caption{ Representation of the the logarithm of the eigenvalues $\lambda_i$ of the reduced density matrix of the system at $\gamma=0.8$. We use OBC and $N=32$ sites on each chain. For each value of $\lambda_i$, it is possible to see horizontally the corresponding degeneration given by the number of blue dots.}
\label{fig:Autovalori08} 
\end{figure}
Finally we analyze the behaviour of the entanglement entropy and the entanglement spectrum. Using again OBC, we calculate the von Neumann entropy $S_v$ and the spectrum of the reduced density matrix \cite{PollmannTurnerBergOshikawa} obtained by tracing out half of the chain.\\
\textcolor{red}{Fig.\ref{fig:Entropia} shows the values of $S_v$ for $\gamma$ from 0 to 1 with a 0.05 step, obtained for a ladder with $N=32$ sites on each chain and OBC. We observe a clear high peak near the $\gamma$-range in which the gap in Fig.\ref{fig:energylevelsOBC} closes. This is in agreement with the fact that, in the thermodynamic limit,  entanglement entropy $S_v$ diverges at the critical point $\gamma_c$. }\\
Also, we know that at least \textcolor{red}{an even} degeneracy of the entanglement spectrum \cite{PollmannTurnerBergOshikawa} is expected in the topological phase. In Fig.\ref{fig:Autovalori02} and Fig.\ref{fig:Autovalori08}, we show the logarithm of the eigenvalues of the reduced density matrix greater than $10^{-12}$, for $\gamma=0.2$ and $\gamma=0.8$ respectively, again obtained for a ladder of $N=32$ sites on each chain. We can summarize our results by noting that the degeneracy of the entanglement spectrum changes from odd for $\gamma=0.2$ (Fig.\ref{fig:Autovalori02}) to even for $\gamma=0.8$ (Fig.\ref{fig:Autovalori08}), in agreement with the fact that we find a non-trivial phase for $\gamma > \gamma_c$. 

\section{Conclusion }
\label{sec:conclusione}

We analyzed a two-legged spin ladder with alternated interactions.\\
Using a path integral formulation of the partition function based on spin coherent states, we analytically mapped the system into a NL$\sigma$M plus a topological term. This allowed us to confirm \cite{MartinShankarSierra,MartinDukelskySierra,NataliaChepiga} that, for a certain value of the parameter $\gamma$ which characterizes the interaction, there is a phase transition. We note here that the numerical result for $\gamma_c$ seems to be very close to half of our theoretical prediction. The same discrepancy was found for spin $1$ chains with staggered interaction \cite{Sierra2,Yamamoto}. This effect may be due to different causes, such as lattice and finite-size effects, perturbative and non-perturbative renormalization corrections to the semiclassical (large $s$) approximation on which the Haldane map is based, an implicit dependence on the number of legs. With our present knowledge we are not able to sort these different effects out, but this does not affect our main findings.\\
We then performed a numerical study based on the DMRG algorithm to characterize the two different gapped phases. In particular we saw that the $\gamma>\gamma_c$  phase is accompanied by a set of zero modes, hinting that it corresponds to an SPT order. This was confirmed by the analysis of the correlations between the spins at the edges and by looking at the degeneracy of the entanglement spectrum. We also calculated some NLOP, showing that the parity and the string order parameters have a dual behaviour, with the former being non zero for $\gamma<\gamma_c$ and the latter for $\gamma>\gamma_c$.\\
Following the classification of \cite{MontorsiRoncaglia,BarbieroMontorsiRoncaglia,MontorsiDolciniIottiRossi}, we can say that we can identify the region for $\theta<\theta_c$ with a Mott insulator-like phase and the region with $\theta>\theta_c$ with a Haldane insulator-like phase.

In conclusion, \textcolor{red}{our results show that the} presence of a topological term in the NL$\sigma$M induces a critical point which separates an ordinary phase from a topological one.\\
A similar situation\textst{s} may be encountered in other systems. For example, it would be interesting to extend this analysis to ladders with more than two legs, possibly going toward the two-dimensional limit. Also an analogous study could be performed in the case of higher spin $SU(2)$ \cite{Haldane1,Haldane2} or even $SU(N)$ \cite{Affleck1} systems.

\section*{Acknowledgments}
\label{sec:acknowledgments}
G.G., G.M. and E.E. are partially supported through the project "ALMAIDEA" by University of Bologna. G.M. and E.E. are partially supported through the project "QUANTUM" by Istituto Nazionale di Fisica Nucleare (INFN).\\
\\

\textcolor{red}{
\section*{Appendix}}
\textcolor{red}{In this appendix we give some details about the numerical analysis and in particular the finite-size scaling procedure we adopt.}\\
\textcolor{red}{We perform the numerical analysis by means of a finite-size DMRG code which is based on tensor networks (TN). ITensor library \cite{ITensor} allows to describe the wave function of a system by using a network of interconnected tensors. In this framework, matrix product states (MPS) are TN states that correspond to a one-dimensional array of tensors \cite{ver,sch,TN}. \textcolor{red}{In a similar way, operators as the Hamiltonian are represented by means of matrix product operators (MPO), yielding powerful tools to describe a wide range of models including one dimensional chains and ladders. }TN codes work better with open boundary conditions (OBC) than with periodic boundary conditions (PBC). However, if it is necessary to analyze both boundary conditions, as we do, PBC can be easily achieved from OBC code adding the appropriate interaction term between the first and last site of each chain in the Hamiltonian.\\
Since ITensor Network Library does not implement the conservation of the total spin quantum number, our simulations \textcolor{red}{are implemented in the whole Hilbert space and we target the ground state and the first three/four excited states.} This multi-target approach allows us to check that the $SU(2)$ symmetry is respected also at the level of numerical simulations, a fact which is also confirmed by the behavior of NLOP (Fig.\ref{fig:NLOPwithPBC} and Fig.\ref{fig:NLOPwithOBC}) and spin correlation functions (Fig.\ref{fig:Corr015alternata}, Fig.\ref{fig:Corr015} and Fig.\ref{fig:Corr1}).\\
We work with up to $N$ sites for each chain, $N=30$ for PBC and $N=32$ for OBC; \textcolor{red}{in the calculations of the energy spectrum we usually consider 7 sweeps for each state, while} for spin correlation functions we consider 40 sweeps. For each sweep, the maximum value of bond dimension varies from 10 to 500, the minimum value of bond dimension varies from 10 to 20. \textcolor{red}{Also, the cutoff changes from $10^{-5}$ to $10^{-10}$ and the noise term, which is added to the density matrix to help convergence, varies from $10^{-5}$ to $0$.}}\\
\begin{figure}
\centering
{\includegraphics[width=0.40\textwidth]{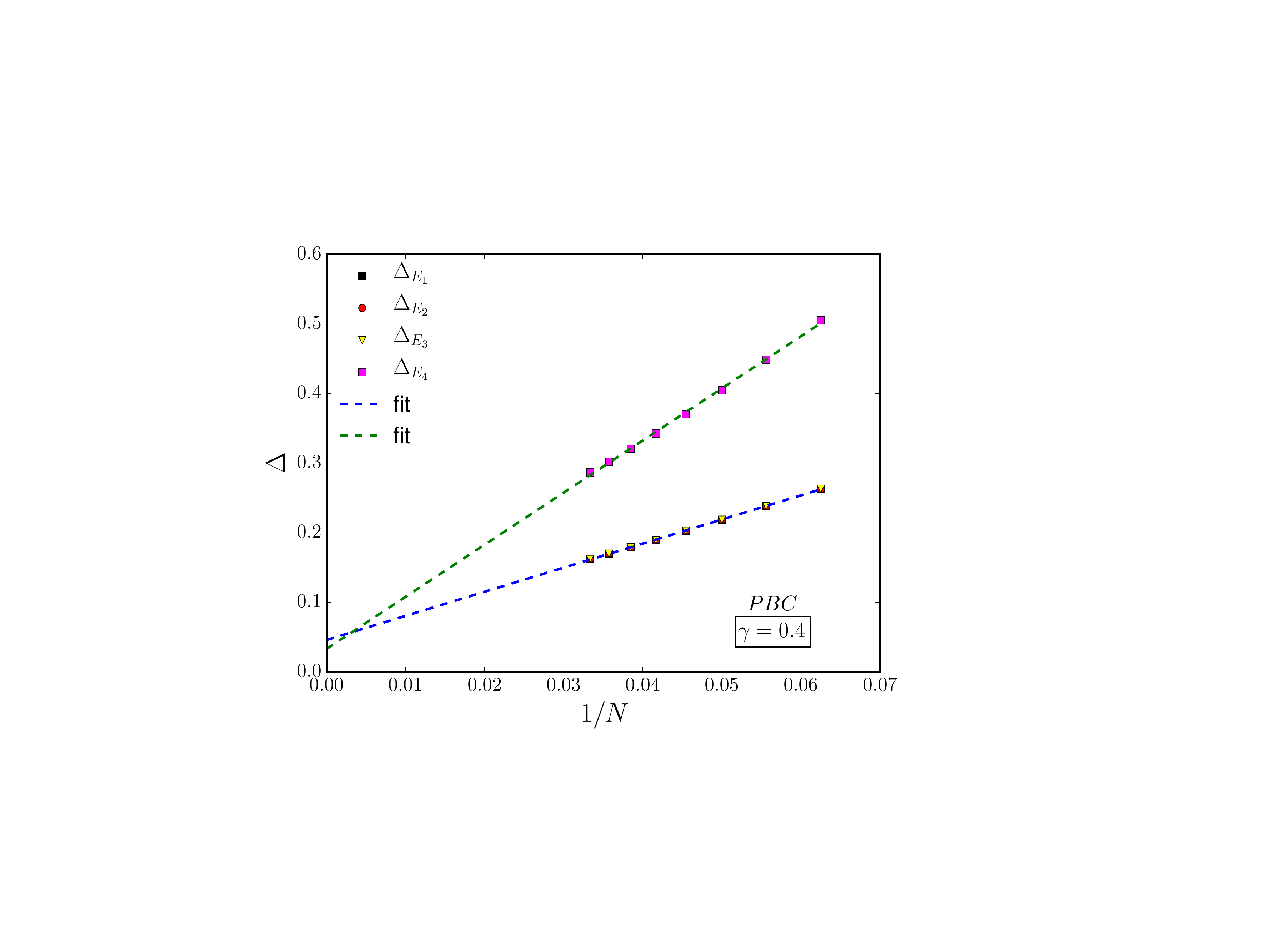}}
\hspace{2mm}
{\includegraphics[width=0.4\textwidth]{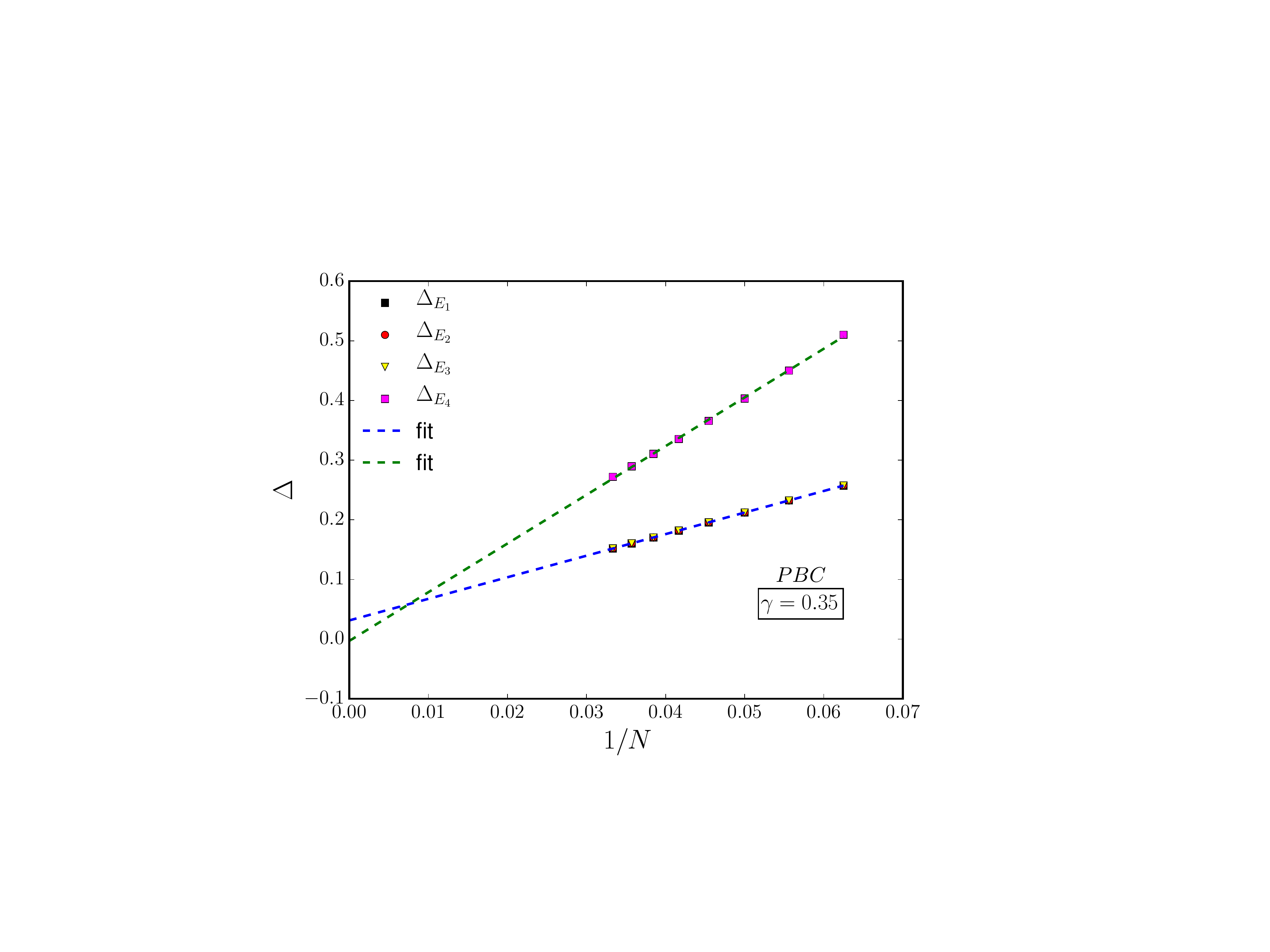}}
{\includegraphics[width=0.4\textwidth]{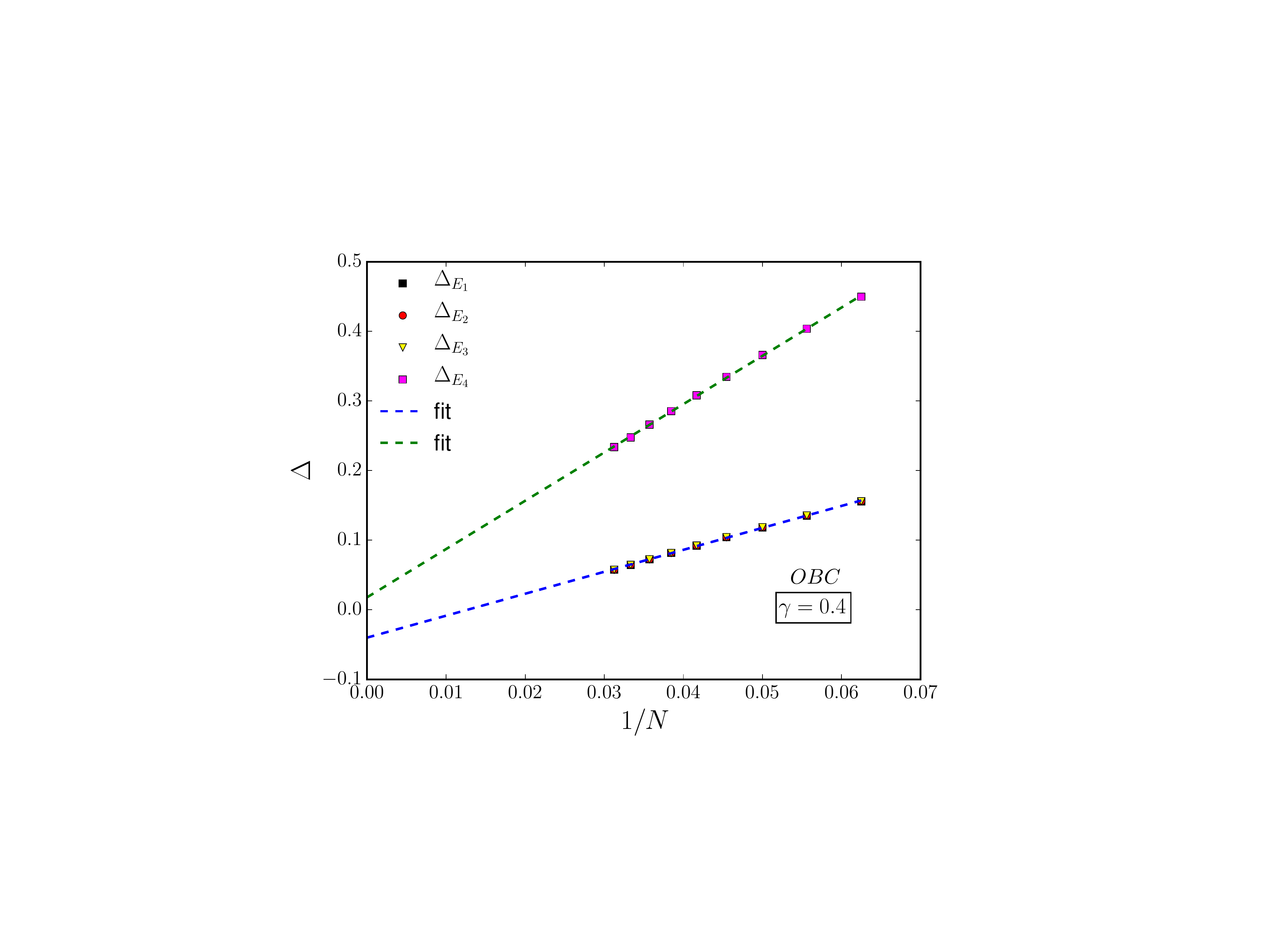}}
{\includegraphics[width=0.40\textwidth]{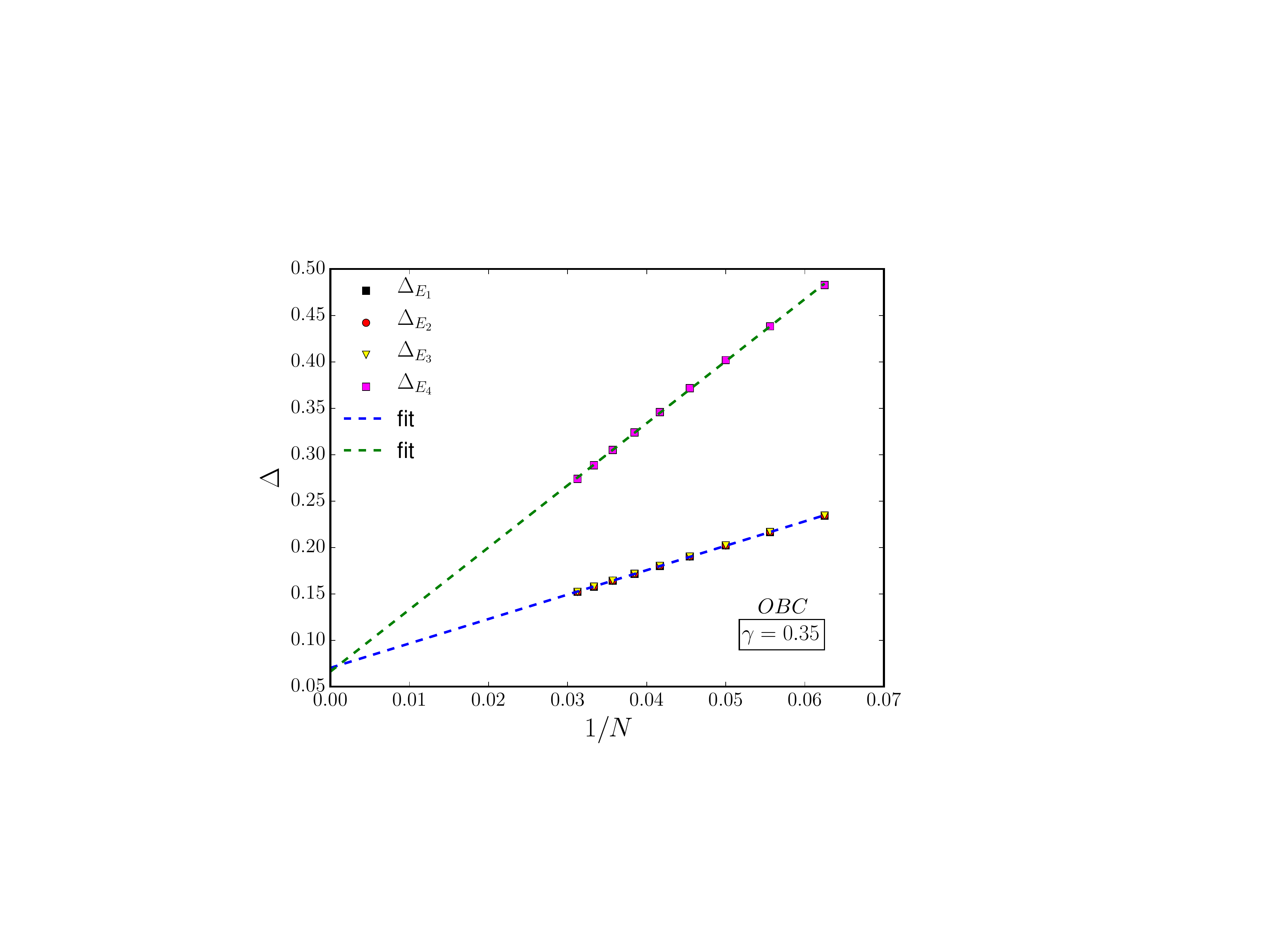}}
\caption{ \textcolor{red}{Scaling of the energy gap of each state of the triplet $\Delta_{E_1}$, $\Delta_{E_2}$, $\Delta_{E_3}$ and of the energy gap of the fourth excited state $\Delta_{E_4}$ with respect to the ground state; we consider for PBC $N=16, 18, 20, 22, 24, 26, 28, 30$ sites on each chain and for OBC $N=16, 18, 20, 22, 24, 26, 28, 30, 32$ sites on each chain. First panel: PBC and $\gamma=0.4$. Second panel: PBC and $\gamma=0.35$. Third panel: OBC and $\gamma=0.4$. Fourth panel: OBC and $\gamma=0.35$.}}
\label{fig:nuovo_scaling}
\end{figure}

\begin{figure}
\centering
{\includegraphics[width=0.55\textwidth]{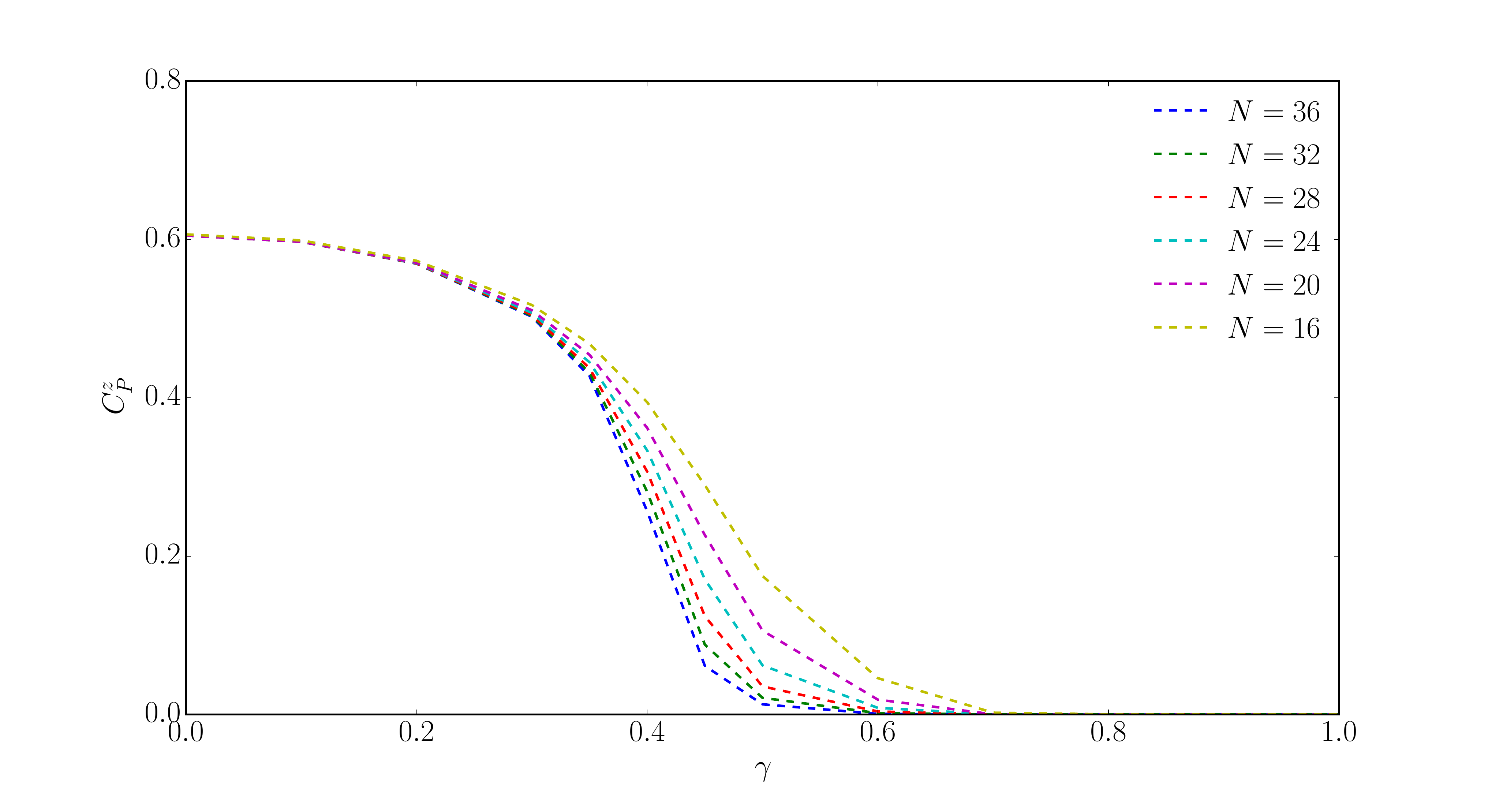}}
{\includegraphics[width=0.55\textwidth]{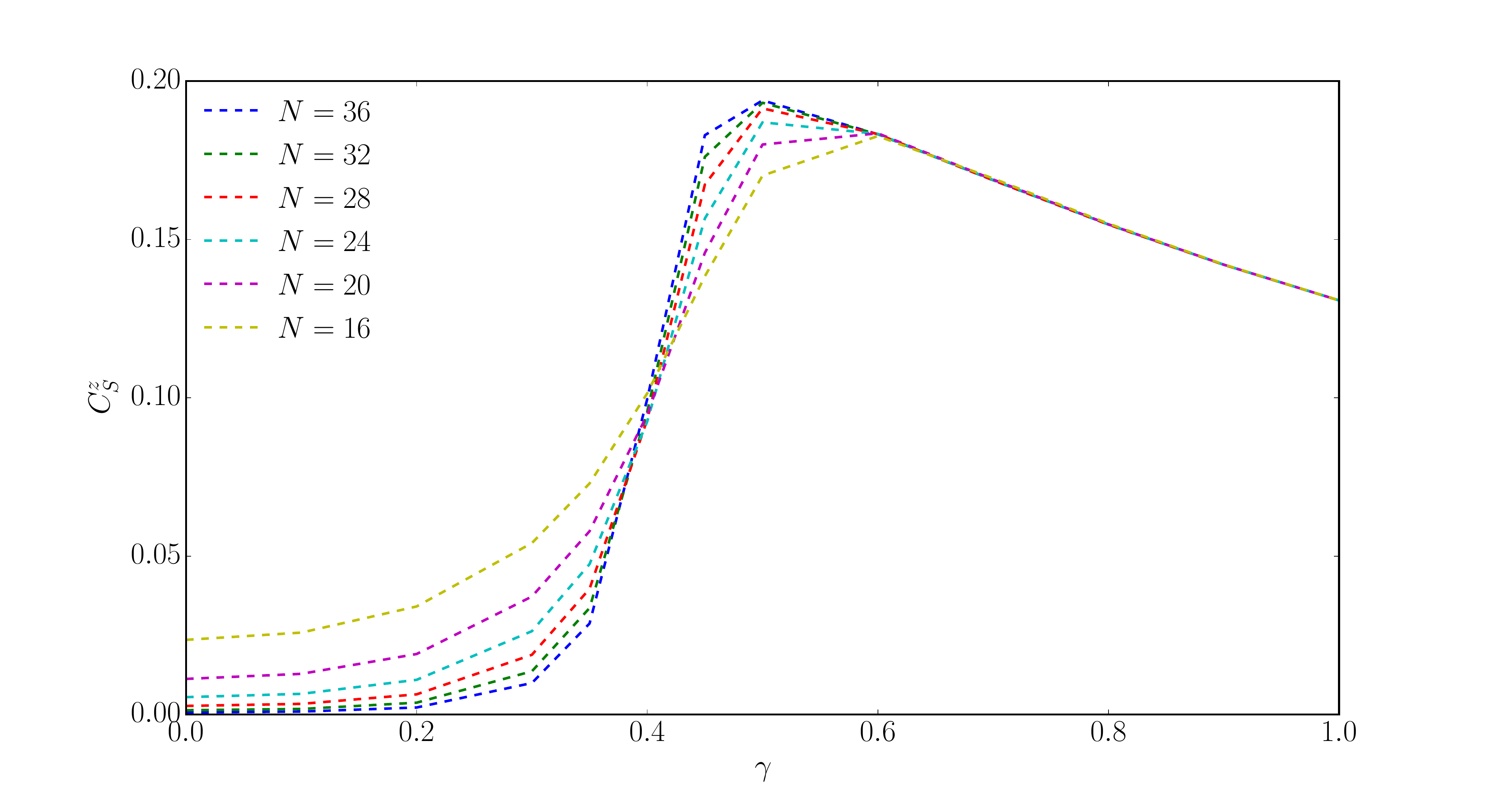}}
{\includegraphics[width=0.55\textwidth]{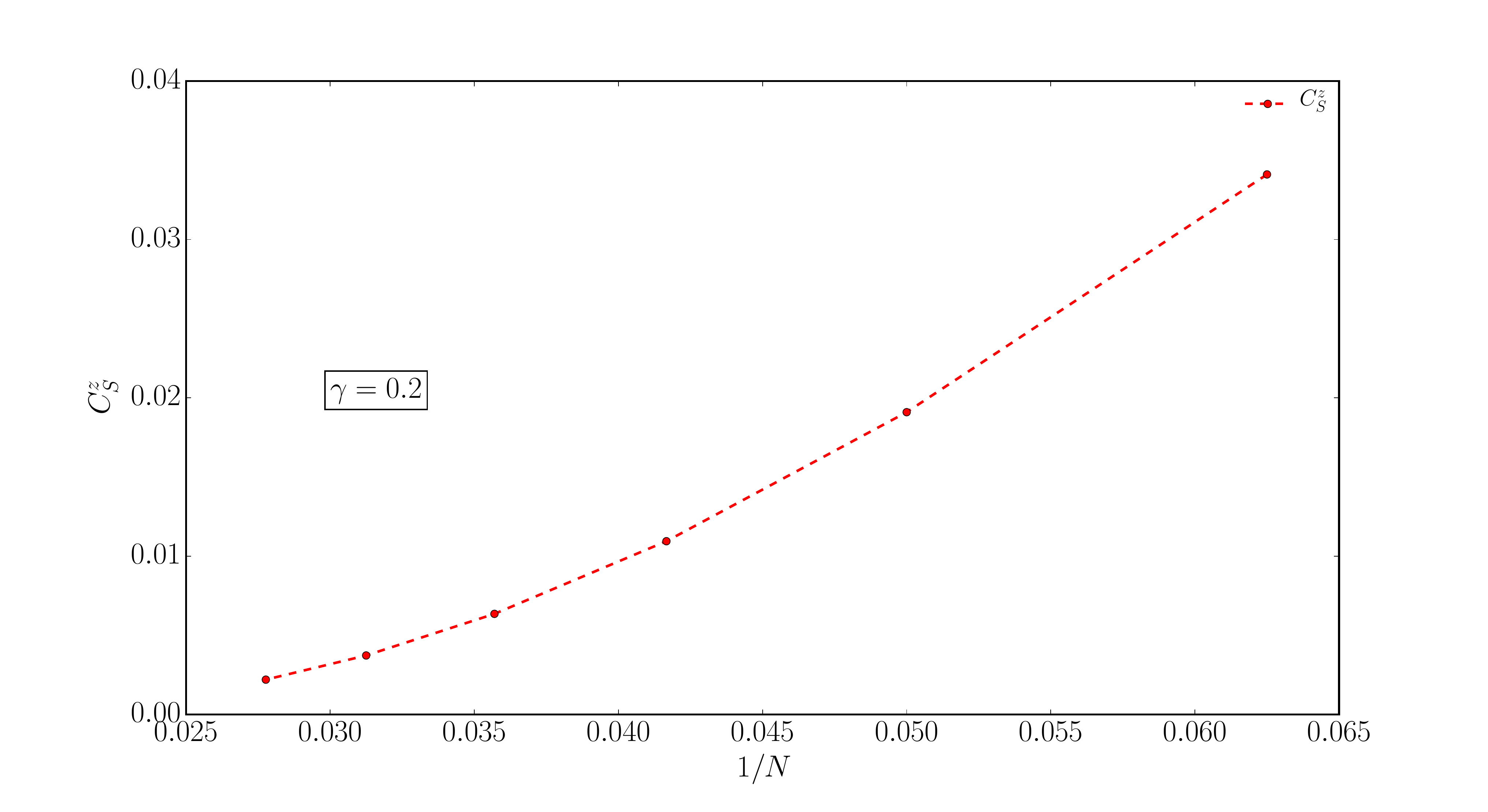}}
{\includegraphics[width=0.55\textwidth]{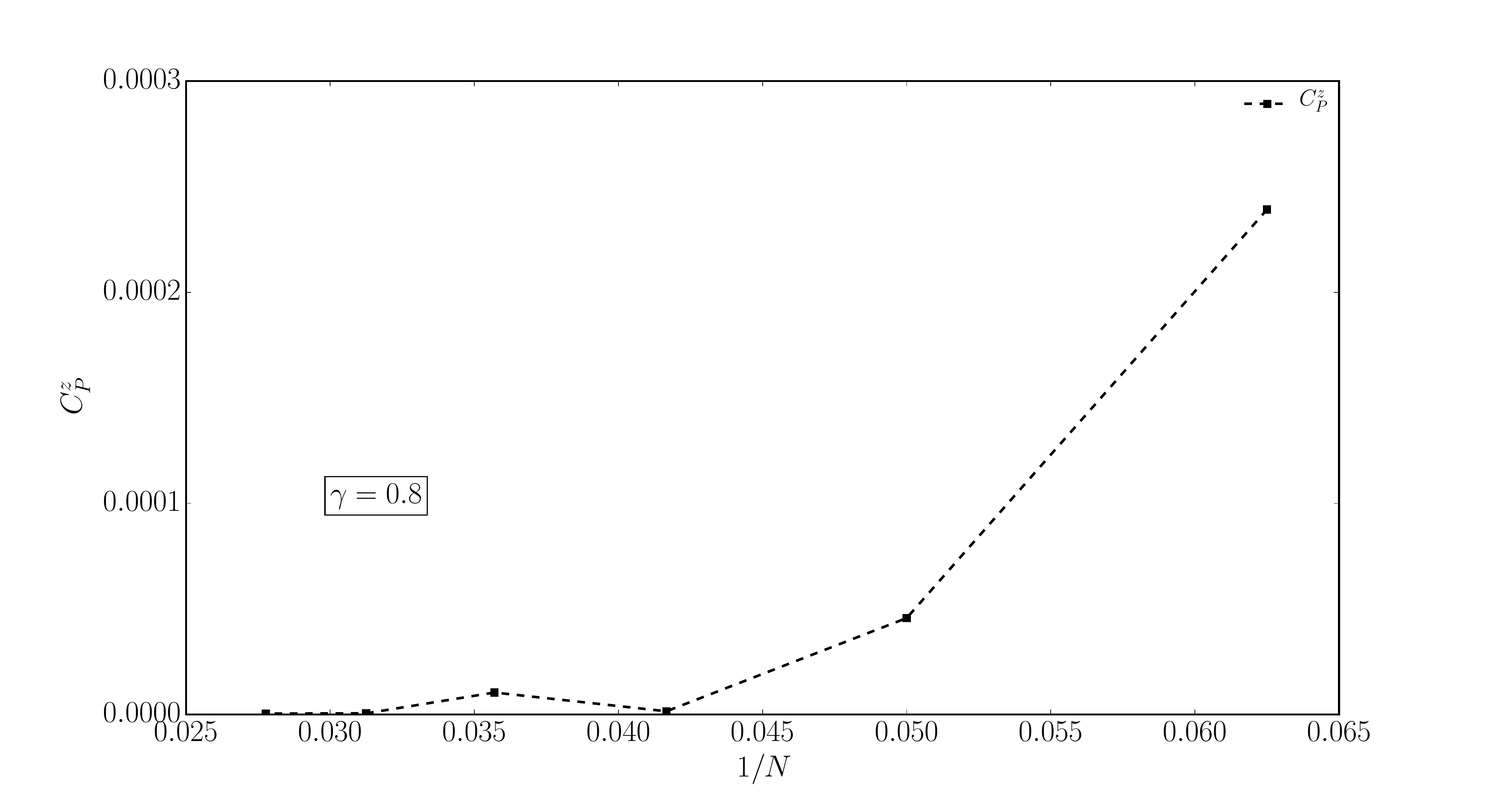}}
\caption{ \textcolor{red}{Finite size-scaling of NLOP.  First and second panel: parity $C_P^z$  and string $C_S^z$ order parameters respectively, as function of $\gamma$,  for OBC and different sizes. 
Third panel: string order parameter $C_S^z$, as function of $1/N$ in the trivial phase with $\gamma  = 0.2$.
Fourth panel: parity order parameter $C_P^z$, as function of $1/N$ in the topological phase with $\gamma  = 0.8$.}\\}
\label{fig:NLOP_confronto_OBC} 
\end{figure}

\textcolor{red}{All the physical quantities that we have evaluated (energy gap, NLOP, spin correlation functions, entanglement entropy and spectrum) are consistent with the analytical prediction of a critical point $\gamma_c$ and the emergence of two different phases. As already anticipated in section \ref{sec:numerico}.A, our aim is not the exact determination of the critical point $\gamma_c$, but the analysis of the two gapped phases, to show that the one for $\gamma>\gamma_c$ is topological. For completeness, however, here we present some graphs describing the finite-size behaviour of the different physical quantities of interest.\\
First, we consider the scaling of the first four excited energy states close to the critical point. This is shown in Fig.\ref{fig:nuovo_scaling} for both PBC and OBC, in the cases of $\gamma=0.35$ and $\gamma=0.4$. These data clearly indicate that we are very close to a phase transition point, whose accurate location would however require bigger sizes of the ladder in order to have results which are  independent of boundary conditions. \\
Second, we describe how the the NLOPs scale with the size of the system. In particular, since the $SU(2)$ symmetry is respected, in the first two panels of Fig.\ref{fig:NLOP_confronto_OBC} we only report the values of the $z$-component of the parity and string order parameters as functions of $\gamma$, for different system sizes ($N=16,20,24,28,32,36$) and OBC.  Then, in the third panel we show how the string order parameter scales to zero in the trivial phase, for $\gamma = 0.2$, while in the fourth panel we show how the parity order parameter scales to zero in the topological phase, for $\gamma = 0.8$.\\
All these data confirm that the maximum size at which we perform the numerical simulations ($N =30$ for PBC and $N =32$ for OBC) is large enough to ensure small errors and accuracy of the results.}

\end{document}